\documentclass[preprintnumbers,amsmath,amssymb]{revtex4}

\usepackage{graphicx}
\usepackage{dcolumn}
\usepackage{bm}

\newcommand{\beq}{\begin{equation}}
\newcommand{\eeq}{\end{equation}}
\newcommand{\bdis}{\begin{displaymath}}
\newcommand{\edis}{\end{displaymath}}
\newcommand{\bea}{\begin{eqnarray}}
\newcommand{\eea}{\end{eqnarray}}
\newcommand{\barr}{\begin{array}}
\newcommand{\earr}{\end{array}}
\newcommand{\bfig}{\begin{figure}[!]}
\newcommand{\efig}{\end{figure}}

\begin{document}


\title{Basics of the magnetocaloric effect}

\author{Vittorio Basso}
\affiliation{Istituto Nazionale di Ricerca Metrologica, Strada delle Cacce 91, 10135 Torino, Italy}

\date{April, 7, 2013}

\begin{abstract}
This chapter reviews the basic physics and thermodynamics that govern magnetocaloric materials. The thermodynamics of magnetic materials is discussed by introducing relevant free energy terms together with their microscopic origin leading to a discussion of the sources of entropy that can change in an applied magnetic field.  Such entropies account for measurable magnetocaloric effects, especially in the vicinity of magnetic phase transitions.  Particular attention is devoted to first order magnetic transitions that involve the coupling of spin, lattice, electronic and anisotropic magneto-crystalline degrees of freedom. The problem of irreversibility and hysteresis, present in magnetocaloric materials with first order phase transitions is discussed in the context of out-of-equilibrium thermodynamics and hysteresis modeling.
\end{abstract}

\maketitle

\tableofcontents
\section{Introduction}

The recent discoveries of a so-called ``giant'' magnetocaloric effect (MCE) in alloys such as Gd$_5$Si$_2$Ge$_2$, La(Fe,Si)$_{13}$ and (Mn,Fe)$_2$(P,Z), have driven strong research efforts focused on its maximization.   A principal aim has been the development of magnetic cooling devices working around room temperature, which will use such alloys as their refrigerant.  In this chapter, we review the basics of the magnetocaloric effect by first considering the general physics of magnetic materials \cite{Chikazumi-1964, Morrish-1965, Herpin-1968, Bozorth-1993, Bertotti-1998, OHandley-2000, Coey-2009} and the relevant thermodynamics \cite{Wilson-1953, Wannier-1966, Callen-1985, Landau-1986} that governs magnetocaloric properties \cite{Tishin-2003, Gschneidner-2005, Tishin-2007, Bruck-2008, Gschneidner-2008, Shen-2009, DeOliveira-2010}.  

This chapter is organized as follows. In Section II we introduce the magnetocaloric effect in the context of equilibrium thermodynamics. In Sections III and IV we introduce the thermodynamics of magnetic materials by discussing the microscopic origin of the relevant free energy terms. We are particularly interested to review the mechanisms that give rise to magnetic phase transitions and the sources of a magnetic field-induced entropy change in their vicinity, since the largest MCE is found around such transitions.  Particular attention is devoted to first order magnetic transitions which involve the coupling of different degrees of freedom including: spin, lattice, electronic and magneto-crystalline anisotropy.  

This is an active field of research where different approaches and interpretations have been proposed and are currently widely discussed in the literature \cite{Tishin-2003, DeOliveira-2010}. We introduce the basic ideas underlying different approaches with the aim of presenting their conceptual basis together with their intrinsic limitations. We describe: in Section III the second order magnetic phase transition of a ferromagnet in the mean field theory and in Section IV, the first order phase transition in the Bean-Rodbell model of magneto-elastic coupling \cite{Bean-1962}. In both cases we discuss the consequences for the magnetocaloric effect. In Section V we touch the problem of the irreversibility and hysteresis, because most of the magnetocaloric materials with first order phase transitions display temperature hysteresis as well as magnetic field hysteresis. We  refer to concepts and models that have been developed to describe magnetic hysteresis in ferromagnets and can be extended to phase transitions \cite{Bertotti-2006}.   Through a basic understanding of the mechanisms of first order magnetic phase transitions we hope to fully exploit the cooling potential of magnetic materials and to make magnetic refrigeration at room temperature a viable alternative to conventional refrigeration technologies.

\section{Thermodynamics and magnetocaloric effect}

\subsection{Equilibrium thermodynamics}

\subsubsection{Thermodynamic potentials and equations of state}

Equilibrium thermodynamics, or Gibbs thermostatics, is a theory that applies to systems which are uniquely defined by the values of their state variables \cite{Callen-1985}. It is natural to take as state variables the set of its extensive properties: the internal energy $U$, the volume $V$, the magnetic moment $m$, the entropy $S$ and so on. The system is then defined once a relation connecting the state variables are known. This expression is called the fundamental equation and consists of the expression relating internal energy $U$ to all the other state variables $U(m,V,S,...)$. The corresponding intensive conjugated variables are defined by the derivatives of the internal energy.  For example: the pressure is $p = -\partial U/\partial V$, the magnetic field is $\mu_0H = \partial U/\partial m$, where $\mu_0$ is the permeability of free space, the temperature is $T = \partial U/\partial S$, and so on. When dealing with material properties it is useful to introduce specific quantities as volume densities or mass densities. For solid magnetic materials it is reasonable to assume that mass is conserved and to allow the volume to change. Hence for magnetocaloric materials it is then appropriate to use specific extensive variables, calculated as mass densities. We here introduce: the specific internal energy $u$, the specific volume $v$, the magnetization $M$, the specific entropy $s$ and so on. In order to have an explicit dependence on the conjugated intensive variables rather than on the extensive ones, the specific free energy $f$ and Gibbs free energy $g$ have also to be introduced. The free energy $f(M,v,T)$ depends explicitly on the temperature $T$ and has the following relation with the internal energy $f = u -Ts$. Consequently, the derivatives of the specific free energy $f(M,v,T)$ gives the magnetic field 

\beq
\mu_0 H = \frac{\partial f}{\partial M}
\label{EQ:Hf_def}
\eeq

\noindent the pressure
\beq
p = - \frac{\partial f}{\partial v}
\label{EQ:pf_def}
\eeq
\noindent and the specific entropy

\beq
s = -\frac{\partial f}{\partial T}
\label{EQ:sf_def}
\eeq

\noindent The relations obtained by these derivatives are the state equations. They express the dependence of $H$, $p$ and $s$ on the independent variables $M$, $v$ and $T$. The specific Gibbs free energy $g(H,p,T)$ which is related to the specific free energy by $ g = f - \mu_0HM + pv$ and depends only on the intensive variables. For a magnetic material the specific Gibbs free energy is particularly useful because the intensive variables magnetic field $H$, pressure $p$ the temperature $T$ are often the externally controlled variables in experiments. The derivatives of the Gibbs potential gives the magnetization:

\beq
\mu_0 M = - \frac{\partial g}{\partial H}
\label{EQ:Mg_def}
\eeq

\noindent the volume
\beq
v = \frac{\partial g}{\partial p}
\label{EQ:vg_def}
\eeq

\noindent and the specific entropy

\beq
s = -\frac{\partial g}{\partial T}
\label{EQ:sg_def}
\eeq

\noindent Since the three state equations are obtained by the derivatives of the same function $g(.)$, it turns out that they are not independent of each other. Due to the properties that an equilibrium thermodynamic potential must satisfy \cite{Callen-1985}, the second mixed derivatives of $g(.)$ coincide, with the consequent relations being known as Maxwell relations:

\beq
\left. \mu_0 \frac{\partial M}{\partial T} \right|_{H,p}= \left. \frac{\partial s}{\partial H} \right|_{p,T} 
\label{EQ:MaxwellMs_def}
\eeq

\beq
\left. \frac{\partial v}{\partial T} \right|_{H,p} = - \left. \frac{\partial s}{\partial p}  \right|_{H,T}
\label{EQ:Maxwellvs_def}
\eeq

\beq
\left. \mu_0 \frac{\partial M}{\partial p} \right|_{H,T}= - \left. \frac{\partial v}{\partial H} \right|_{p,T}
\label{EQ:MaxwellMv_def}
\eeq

\subsubsection{Demagnetizing effects}

When the previous definitions are extended to vector quantities, each component of the magnetic field vector is given by the derivative with respect to the relative magnetization component: $\mu_0H_x = \partial u/\partial M_x$ and so on. If the magnetic system consists of a body of finite size, we have also to take explicitly into account the energy term associated with the creation of a magnetostatic field $\vec{H}_M$ generated by the magnetization distribution in space \cite{Herpin-1968, Bertotti-1998}. $\vec{H}_M$ is given by the solution of the magnetostatic Maxwell equations $\nabla \cdot \vec{H}_M= - \nabla \cdot \vec{M}_v$ and $\nabla \times \vec{H}_M = 0$ and the energy of the magnetostatic field is given by the integral extending over the magnetic body volume $V$:

\beq
U_M = - \frac{\mu_0}{2}\int_V \vec{H}_M \cdot \vec{M_v} \, d^3r
\label{EQ:UM_def}
\eeq

\noindent where $\vec{M}_v$ is the magnetization vector as volume density. The magnetostatic energy depends on the internal distribution of the magnetization, however when it is reasonable to consider the magnetization as uniform inside the body, the problem is greatly simplified because the  magnetostatic field is due to the distribution of the magnetization at the sample surface only. A simplifying case is when the sample is ellipsoidal; then the magnetostatic field is spatially uniform inside the body. By taking the reference frame along the tree main axis $(a,b,c)$ of the ellipsoid, the magnetostatic energy is 

\beq
U_M =  V \frac{1}{2} \mu_0\left(  N_aM_{v_x}^2+N_bM_{v_x}^2+N_cM_{v_z}^2 \right)
\label{EQ:UMdem_def}
\eeq

\noindent where the dimensionless proportionality factors are the demagnetizing coefficients which depend only on the aspect ratios of the ellipsoid and have the property $N_a+N_b+N_c=1$. In the case of spatial uniformity the magnetostatic field is also called demagnetising field $\vec{H}_d = \vec{H}_M$, because, as it can be seen by taking the derivatives of  Eq.\ref{EQ:UMdem_def}, it is proportional to the magnetization components, but oriented in the opposite direction $\vec{H_d} = - ( N_a M_{v_x} + N_b M_{v_y} +N_c M_{v_z}$. In presence of both an applied field (i.e. applied by suitable coils) $\vec{H}_a$ and the demagnetizing field $\vec{H}_d$, the two contributions superpose to give the magnetic field $\vec{H}$: $\vec{H} = \vec{H}_a+\vec{H}_d$. 

One of the problems of the thermodynamics of magnetism is if the magnetostatic energy has to be included in the thermodynamic internal energy $U$ or not \cite{Herpin-1968, Bertotti-1998}. As often in thermodynamics, the choice is left to the analysis of the experimental constraints. In fact the result is that, if the magnetostatic energy term is not included in $U$, then the intensive variable coupled to the magnetization is the field $\vec{H}$, while if it is included in $U$, the intensive variable coupled to the magnetization results to be the applied field $\vec{H}_a$. In many experimental situations, it may be relatively easy to control the applied magnetic field $\vec{H}_a$, while the control of $H$ may require a detailed knowledge of the demagnetizing coefficients and a feedback control on the sources of $\vec{H}_a$. If the applied field $\vec{H}_a$ is used a field variable, all the thermodynamic relations derived in the previous section are still valid, but one has to bear in mind that the internal energy of the system and all the thermodynamic potentials will contain also the energy of the demagnetizing field. This mans that the corresponding thermodynamics will depend on the shape of the sample through its aspect ratio.

\subsection{Magnetocaloric effect}

\subsubsection{$\Delta s_{iso}$ and $\Delta T_{ad}$}

\begin{figure}[htb]
\narrowtext 
\centering
\includegraphics[width=18cm]{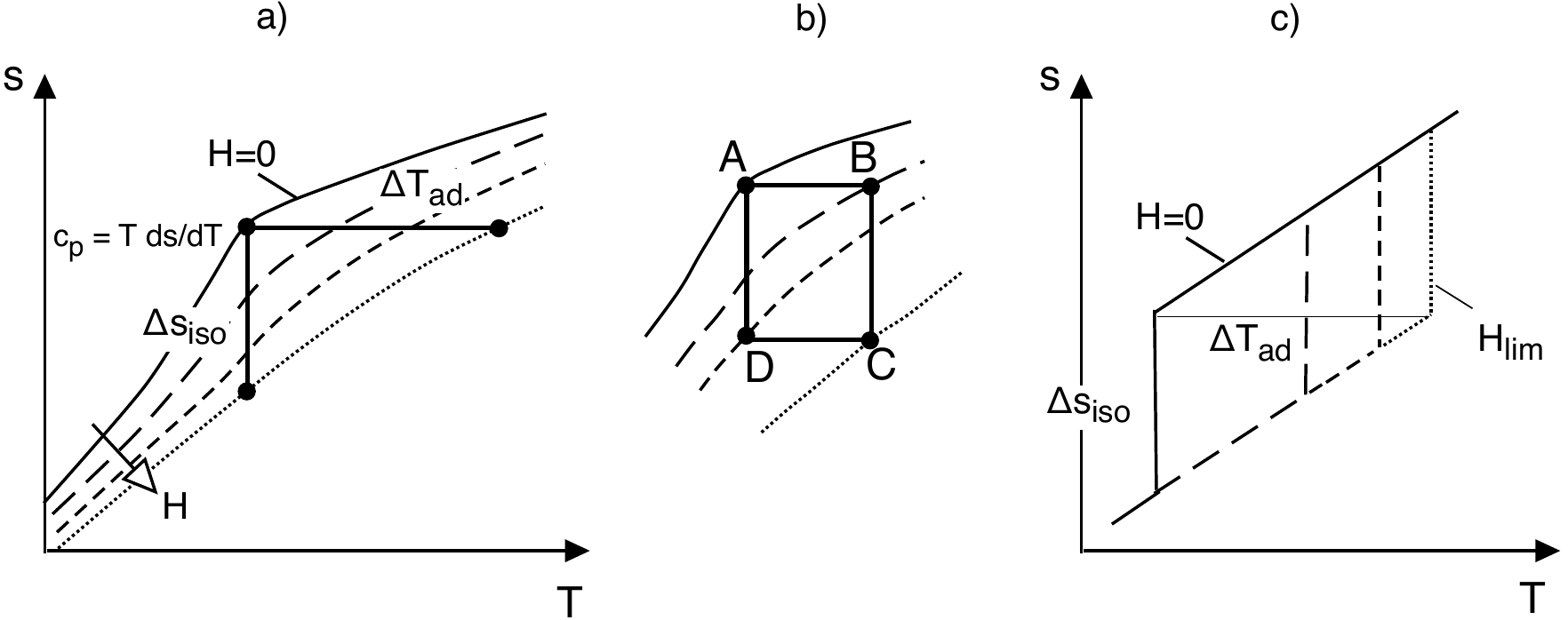}
\caption{a) Entropy state equation $s(H,T)$. b)  Magnetic Carnot cycle in the $(s,T)$ diagram. c) Entropy state equation $s(H,T)$ for an ideal first order transition.} \label{FIG:E2}
\end{figure}

The magnetocaloric effect is defined as the adiabatic temperature change $\Delta T_{ad}$ or the isothermal entropy change $\Delta s_{iso}$ due to the application of the magnetic field $H$ at constant pressure \cite{Tishin-2003}. For systems in thermodynamic equilibrium, the two quantities are derived by the entropy state equation $s(H,T)$ at constant pressure as shown in the sketch of Fig.\ref{FIG:E2}a. The isothermal entropy change $\Delta s_{iso}$ is the difference between two curves at the same temperature, $T$:

\beq
\Delta s_{iso}(H,T) = s(H,T)- s(0,T)
\label{EQ:dsiso_def}
\eeq

\noindent while the adiabatic temperature change is the difference between two curves at the same entropy $s$ (Fig.\ref{FIG:E2}):

\beq
\Delta T_{ad}(H,s) = T(H,s)- T(0,s)
\label{EQ:dTad_def}
\eeq

\noindent The $\Delta T_{ad}$ can also be expressed as a function of the temperature $T=T(0,s)$ at zero magnetic field, giving $\Delta T_{ad}(H,T)$, as it is commonly done in experiments. The two quantities $\Delta s_{iso}(H,T)$ and $\Delta T_{ad}(H,T)$ are not independent because they are related to the slope of the $s(H,T)$ curve as a function of $T$ and therefore to the specific heat:

\beq
c_p(H,T) = T \left. \frac{\partial s}{\partial T} \right|_{H,p}
\label{EQ:cp_def}
\eeq

\begin{figure}[htb]
\centering
\includegraphics[width=18cm]{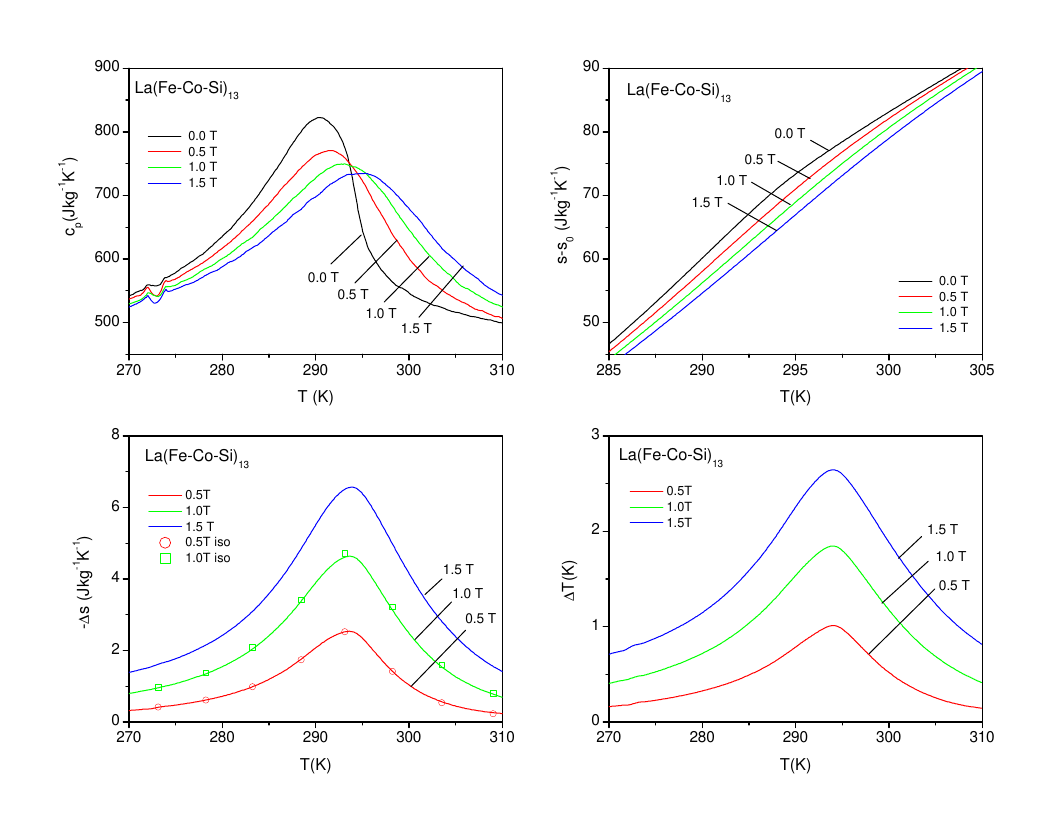}
\caption{Magnetocaloric effect in La(Fe$_{1-x-y}$Co$_y$Si$_x$)$_{13}$ with $x = 0.077$ and $y = 0.079$ \cite{Katter-2008}. Top left: $c_p(H,T)$; top right: $s(H,T)$; bottom left: $\Delta s_{iso}(H,T)$; bottom right: $\Delta T(H,T)$. After Ref.\cite{Basso-2010b}.}
\label{FIG:LaFeCoSi}
\end{figure}

\noindent Magnetic refrigeration cycles can be drawn in the $(s,T)$ diagram as shown in Fig.\ref{FIG:E2}b. Without going into the details of the specific magnetic thermodynamic cycles employed in magnetic refrigeration (see \cite{Kitanovski-2006} and other chapters in this book), we simply observe that cycles of high cooling power and large temperature span can be realized by the maximization of both $\Delta s_{iso}$ and $\Delta T_{ad}$ of the magnetic material. For example in a magnetic Carnot cycle ABCD, $Q_c = T_c\Delta s_{DA}$ is the heat extracted from the cold bath and $\Delta T_{AB}$ is the difference between the hot and cold bath temperatures $T_h - T_c = \Delta T_{AB}$. Such quantities can be derived from the entropy state equation $s(H,T)$ of the magnetic material which in turn can be constructed by the integration of the magnetic field-dependent experimental specific heat \cite{Pecharsky-1999, Pecharsky-2001, Palacios-2005}. An example of the inter-relation of magnetocaloric properties is shown in Fig.\ref{FIG:LaFeCoSi} for La(Fe-Co-Si)$_{13}$.

The Maxwell relation of Eq.(\ref{EQ:MaxwellMs_def}) is particularly important for the MCE, because it relates the entropy $s(H,T)$ equations of state with the the magnetization $M(H,T)$ (all at the same constant pressure). The magnetic field-induced entropy change $s(H_1)-s(H_0)$ can be often be computed from magnetization measurements:

\beq
s(H_1)-s(H_0) = \mu_0 \int_{H_0}^{H_1} \frac{\partial M}{\partial T} dH
\label{EQ:dsMaxwell_def}
\eeq

\noindent and the temperature-induced magnetization change $M(T_1)-M(T_0)$ can be computed from entropy change measurements:

\beq
M(T_1)-M(T_0) = \frac{1}{\mu_0} \int_{T_0}^{T_1} \frac{\partial s}{\partial H} dT \, .
\label{EQ:dMMaxwell_def}
\eeq

\noindent Magnetic materials with a second order phase transition where, at the critical temperature, the magnetic system passes from an ordered ferromagnetic state to a disordered paramagnetic state can be considered to be always in thermodynamic equilibrium. Then the equilibrium relations derived in the previous section apply well. Equation (\ref{EQ:dsMaxwell_def}) gives a practical way to determine the entropy change without the need of calorimetric setups \cite{Pecharsky-1999, Pecharsky-1999b, Palacios-2010}, while Eq.(\ref{EQ:dMMaxwell_def}) has been used in the past to arrive at an accurate determination of the saturation magnetization close to teh critical temperature of magnetic materials \cite{Sucksmith-1953}. Refs.\cite{Pecharsky-1999, Pecharsky-2001} show that in magnetic materials with second order transitions the entropy change $\Delta s(H,T)$ constructed experimentally by either the integration of $c_p(H,T)$ or by the Maxwell relation Eq.(\ref{EQ:dsMaxwell_def}) are in good agreement as expected.

\subsubsection{Thermodynamics of first order phase transitions}

\begin{figure}[htb]
\narrowtext 
\centering
\includegraphics[width=17cm]{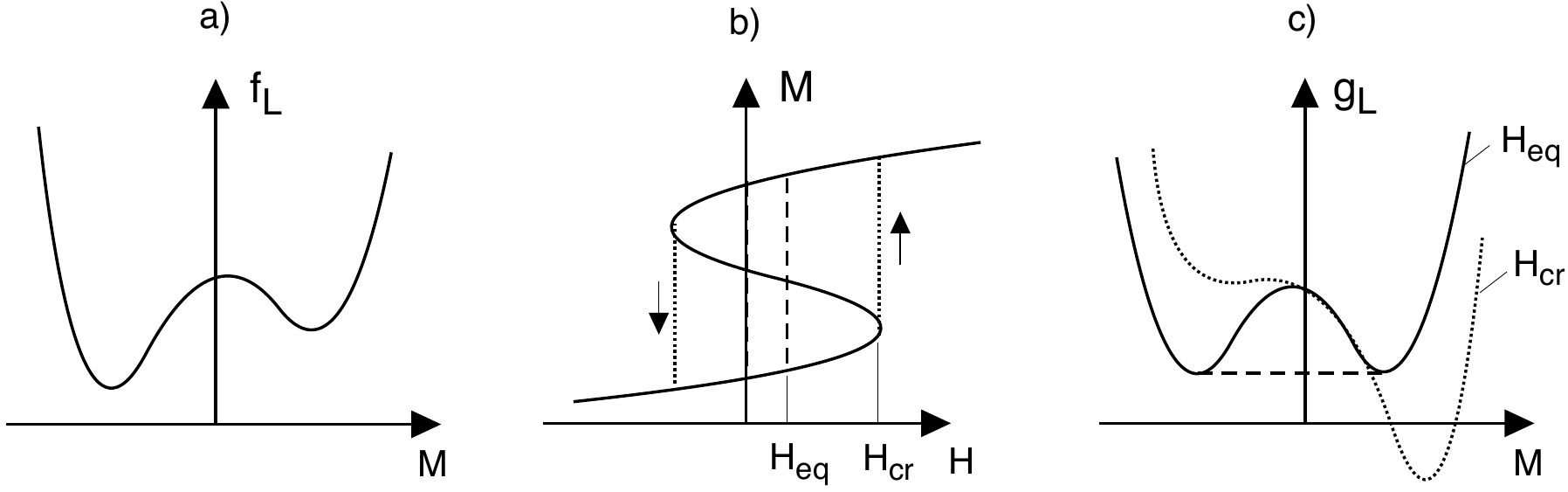}
\caption{a) Free energy potential $f_L(M)$ with two minima. b) Magnetic field state equation $M(H)$. c) Gibbs free energy $g_L = f_L(M) -\mu_0HM$} \label{FIG:fL}
\end{figure}

The equilibrium thermodynamics developed so far requires that the system state corresponds to a global potential energy minimum. When this is not true, the free energy has more then one global minimum, leading to a first order phase transition~\cite{Callen-1985}. In Fig.~\ref{FIG:fL}a, a free energy potential $f_L$ with two minima as a function of the magnetization $M$ is shown as an example. Here we use the subscript $L$ (Landau) to denote that the potential is a non-equilibrium one. When computing the magnetic field state equation $\mu_0H = \partial f_L/\partial M$ corresponding to this example potential, one finds that $M(H)$ has an s-shaped curve (Fig.~\ref{FIG:fL}b). If the magnetic field $H$ is used as controlling variable, there are multiple values of $M$ corresponding to the same $H$, a result which is not compatible with the assumptions made for uniquely defined, equilibrium states. The thermodynamic behavior of such a system characterized therefore has an intrinsically out-of-equilibrium character in the s-shaped region.

Here we are interested on how the system may pass from one minimum to the other by making an abrupt phase transition since such transitions are associated with the largest single changes in entropy and temperature. If one considers the local stability of the energy minima, the evolution of the system state follows a global instability corresponding to the dashed lines of Fig.\ref{FIG:fL}b.  There are two contrasting cases.  The first is the completely out-of-equilibrium picture in which the system transforms into in the new state only at a critical field $H=H_{cr}$ at which the original minimum is completely unstable.  Such a situation also generates hysteresis and is generally followed only if there are no other energetically favorable ways to pass to the low energy minimum before the instability occurs. However, the macroscopic system always possesses many internal degrees of freedom by which, with the contribution of spontaneous fluctuations, they are generally able to transform to the new phase before the global instability. From this idea of a phase transition a second situation arises when it is possible to use the Maxwell convention in which the system may spontaneously select the minimum with the lowest global Gibbs free energy $g_L = f_L -\mu_0HM$ (Fig.\ref{FIG:fL}c). If so, then equilibrium behavior is recovered as the selection of the lowest minimum has the effect to remove the effects of the energy barrier. At the field $H=H_{eq}$ when the two minima have the same energy level (dashed line of Fig.\ref{FIG:fL} right) the system can be indifferently in one phase or the other or in a phase coexistence state at no additional energy cost. The corresponding phase transformation (Fig.\ref{FIG:fL}b) is a vertical line without hysteresis. 

A limit case which is of interest for magnetic refrigeration is the state equation $s(H,T)$ for an ideal first order equilibrium phase transition in which the entropy has discontinuous change (see Fig.\ref{FIG:E2}c). The temperature at which the transition occurs depends on the magnetic field $H$ and its derivative is given by the Clausius-Clapeyron equation:

\beq
\frac{dT}{dH} = - \mu_0 \frac{\Delta M}{\Delta s}
\label{EQ:CC_def}
\eeq

\noindent where $\Delta M$ and $\Delta s$ are the discontinuous changes of the magnetization and the entropy at the transition. If $\Delta M$ and $\Delta s$ are constant values, then one obtains that, for a magnetic field variation from 0 to $H$, the transition temperature changes by an amount $\Delta T = \mu_0 H {\Delta M}/{\Delta s}$. The adiabatic temperature change is limited by the specific heat value. By taking the ratio $(c_p/T)$ as a constant value, the upper limit is $\Delta T_{ad} = \Delta s/(c_p/T)$ as can be seen from Fig.~\ref{FIG:E2}a.  The energy product $\Delta s_{iso} \cdot \Delta T_{ad}$ is equal to $\mu_0H \Delta M$ if $H<H_{lim}$ and to $(\Delta s)^{2}/(c_p/T)$ if $H>H_{lim}$ with $\mu_0H_{lim}= (\Delta s)^{2}/(\Delta M c_p/T)$. From these simplified relations, one obtains that in an ideal first transition, at a given magnetic field $H$, the maximization of $\Delta M$ gives the maximum energy product, while the ratio of the $\Delta s$ and $(c_p/T)$ determines the upper limit of the adiabatic temperature change~\cite{Zverev-2010,Sandeman-2012}.

\section{Second order magnetic transitions}

As we have seen in Section II, the thermodynamics of a magnetic material can be fully determined by performing experiments. However, it is useful to understand the microscopic physical mechanisms that are at the origin of the magneto-thermal behavior. While the thermodynamics of solids, with the main aim of the prediction of the specific heat, is a well developed classical branch of solid state physics \cite{Wilson-1953,Wannier-1966}, the thermodynamics of magnetic solids, with the aim of the prediction of the magnetocaloric effect, has been the subject of detailed studies only in recent years~\cite{Tishin-2003, DeOliveira-2010}.

While many of the theoretical predictions of electronic structure, the formation of atomic magnetic moments, exchange interactions, the kind of magnetic order and so on, are now based on first principle calculations \cite{Coey-2009}, the thermodynamics of the magnetocaloric effect rely on statistical mechanics~\cite{DeOliveira-2010}. The reason is that the free energy of a magnetic material is the consequence of the presence of several contributions to the system entropy including: the atomic magnetic moments (due to electron spin and orbital momentum), the lattice vibrations and the electronic states.

\subsection{Entropy of magnetic moments}

The effect of the magnetic field on the entropy due to atomic magnetic moments can be appreciated by considering the thermodynamics of an ensemble of non interaction magnetic moments that give rise to paramagnetic behavior~\cite{Chikazumi-1964, Morrish-1965, Herpin-1968, OHandley-2000}.  We now examine in detail the statistical mechanics of an an ensemble of magnetic moments and discuss how how much this simple model may represent real magnetic materials.

\subsubsection{Statistical mechanics of a paramagnet}

We consider a system composed by magnetic moments localized at the atom sites. The atomic magnetic moment is due to the total angular momentum of the electrons and its projection $m$ along the direction of the magnetic field is $m = - g m_J \mu_B$ where $m_J$ is a number that can assume $2J+1$ discrete values between $+J$ and $-J$, while $J$ is the total angular momentum quantum number due to contribution of the orbital and spin momentum and $g$ is the Land\'e $g$-factor. $\mu_B$ is the Bohr magneton $\mu_B ={e \hbar}/{(2 m_e)}$, $e$ is the electron charge, $m_e$ is the electron mass, $\hbar$ is the Planck constant $h$ divided by $2\pi$. Their values are given in Table~\ref{Tab:constants}.

\begin{center}
\begin{table}
\begin{tabular}{|c|c|c|c|}
	\hline
Boltzmann constant & $k_B$            & $1.381 \times 10^{-23}$ & J  K$^{-1}$ \\
Avogadro constant & $N_A$            & $6.022 \times 10^{23}$ & mol$^{-1}$ \\
Planck constant & $h$            & $ 6.626 \times 10^{-34}$ & m$^2$$\,\,$kg$\,\,$ s$^{-1}$ \\
elementary charge &$e$             & $1.602 \times 10^{-19}$ & A$\,\,$ s\\
electron mass & $m_e$             & $9.109 \times 10^{-31}$ & kg\\
Bohr magneton & $\mu_B$         & $9.27 \times 10^{-24}$ & A$\,\,$m$^{2}$\\
 &$k_B/\mu_B$  & 1.49 & T K$^{-1}$\\
 &$k_BN_A$      & 8.31  & J K$^{-1}$ mol$^{-1}$\\
 &$\mu_BN_A$  & 5.58 & Am$^{2}$mol$^{-1}$\\
	\hline
\end{tabular}
\caption{Values of physical constants}
\label{Tab:constants}
\end{table}
\end{center}

\noindent The energy of the magnetic moment in the magnetic field $H$ is 

\beq
E_0 = \mu_0 g m_J \mu_B H
\label{EQ:E_0_par}
\eeq

\noindent where $\mu_0=4\pi \times 10^{-7}$ is the permeability of free space. Being the atomic moment distinguishable, the partition function $Z$ is given by the sum over the states of the Boltzmann weight

\beq
Z = \sum_{m_J=-J}^{+J} \exp\left( - \frac{E_0}{k_BT}\right)
\label{EQ:Z_par}
\eeq

\noindent where $k_B$ is the Boltzmann constant. The specific Gibbs free energy for an ensemble $n$ magnetic moments per unit mass is $g = - n k_B T \ln Z$ and gives 

\beq
g = - n k_B T \left[ \ln \left[ \sinh\left( \frac{2J+1}{2J}x \right) \right] - \ln \left[ \sinh\left( \frac{1}{2J}x \right) \right]\right]
\label{EQ:g_par}
\eeq

\noindent where the variable $x$ is defined as

\beq
x = \frac{\mu_0 g J \mu_BH}{k_BT} \, .
\label{EQ:x_par}
\eeq

\noindent The magnetization is given by Eq.(\ref{EQ:Mg_def}) and is 

\beq
M = M_0 \mathcal{M}_{J}\left(x\right)
\label{EQ:M_par}
\eeq

\noindent where $M_0 = n m_0 = n g J \mu_B $ is the saturation magnetization at $T=0$, and $\mathcal{M}_{J}(x)$ is the Brillouin function:

\beq
\mathcal{M}_{J}(x) = \frac{2J+1}{2J} \coth \left(\frac{2J+1}{2J}x \right) -  \frac{1}{2J} \coth \left(\frac{1}{2J}x \right) \, .
\label{EQ:Mj_par}
\eeq

\noindent The entropy is given by Eq.(\ref{EQ:sg_def}) and is 

\beq
s = n k_B s_J (x)
\label{EQ:s_par}
\eeq

\noindent where $s_J(x)$ is the Brillouin entropy function

\beq
s_J(x) =   \ln \left[ \sinh\left( \frac{2J+1}{2J}x \right)\right]- \ln \left[ \sinh\left( \frac{1}{2J}x \right) \right] - x \mathcal{M}_{J}(x) 
\label{EQ:sj_par}
\eeq

\noindent By expressing $s_J$ as a function of the normalized magnetization $m=M/M_0$, the first terms of the power series expansion are:

\beq
s_J(m) = \left[ \ln(2J+1) - \frac{1}{a_J}\left( \frac{1}{2}m^2+\frac{b_J}{4}m^4 + \mathcal{O}(m^6) \right) \right]
\label{EQ:sj_exp_par}
\eeq

\noindent where

\beq
a_J = \frac{J+1}{3J}
\label{EQ:aj_par}
\eeq

\noindent and

\beq
b_J =  \frac{3}{10} \,\, \frac{[(J+1)^2+J^2]}{(J+1)^2}
\label{EQ:aj_par}
\eeq

\noindent From Eq.(\ref{EQ:sj_exp_par}) one finds that the entropy of the ensemble of magnetic moments has its maximum at $m=0$, and its value is $s(0) = nk_B \ln(2J+1)$, the upper limit for the entropy associated with the atomic magnetic moments with $2J+1$ discrete levels. It is important to notice that the expression for the entropy of the ensemble of magnetic moments derived here is the direct consequence of the discrete number of energy levels of the magnetic moment is a magnetic field and therefore of the electronic origin of the atomic magnetic moment. The thermodynamics of a magnetic moment taken as a classical vector with continuous orientation would lead to unphysical results as shown in Ref.\cite{McMichael-1992}.

\subsubsection{Magnetic moment and electron spin}

\emph{Localised electrons}. A particularly nice example of magnetism due to localised magnetic moments is given by the partial filling of the 4$f$ shell in the rare earth elements. Although the simple atomic model presented in Section I would apply only to isolated atoms, it turns out that several magnetic solid compounds, in which the interaction between the 4$f$ electrons and the surrounding atoms is small, follow theoretical predictions very well~\cite{Henry-1952, Daudin-1982, Coey-2009}. The same occurs for salts containing transition metal elements with 3$d$ electrons. The main difference is that in 3$d$ elements only the spin momentum contributes to the magnetic moment. This occurs because the wavefunctions of 3$d$ electrons are spatially extended and the orbital momentum is said to be quenched, i.e. suppressed, by the presence of the crystal field of the surrounding atoms \cite{OHandley-2000}.

\emph{Non-localised electrons}. The situation is much more complex when the magnetic moment is due to partially delocalised electrons, as for example in ferromagnetic metals with 3$d$ elements. In the case of metals \cite{Wilson-1953}, electrons can travel from one atom to the other and the wavefunctions are not limited to atomic sites. As a consequence the magnetic moment of one atom is not necessarily a multiple of the electron spin $1\over 2$ and there is no simple theory providing an expression for the entropy of magnetic moments. By considering the electrons contributing to the magnetic moment, the correspondent entropy can be approached by two complementary viewpoints. 

From one point of view, the magnetic electrons can be considered as delocalised and filling the appropriate energy bands, and obey Fermi-Dirac statistics. This means that the contribution to the entropy comes from those electrons lying in an energy band of amplitude $k_BT$ around the Fermi level. This way of looking at the spin entropy has been applied to magnetocaloric materials~\cite{DeOliveira-2010}, however one generally expects a small entropy contribution as this entropy is essentially that of the electrons in Pauli paramagnet \cite{OHandley-2000}. The other way to look at the problem is to consider that, based on experimental observations, the atomic magnetic moment of itinerant ferromagnets does not disappear above the Curie point \cite{OHandley-2000} as in a Pauli paramagnet under zero magnetic field. This means that in a ferromagnetic material the magnetic moment, independently of the localised or delocalised nature of the electrons and of the thermal fluctuations, is formed at the atom site. This argument is supported by the fact that the collective wave functions giving rise to parallel alignment of spins are of the spatially anti-symmetric (anti-bonding) type. These anti-bonding wavefunctions are characterised by high probability densities only at the atom site, because the wavefunction changes sign between adjacent atoms. Conversely, the wavefunctions giving rise to anti-parallel alignment of spins are of the spatially symmetric (bonding) type. These bonding states, with widespread wavefunctions, have lower energy with respect to the anti-bonding ones and fill the low levels of the energy band. Therefore they do not essentially contribute to ferromagnetism \cite{OHandley-2000}. By this observation one may associate the magnetic moment to the atom site and be justified in using a Boltzmann-Gibbs statistical weight for counting the spin states rather then the Fermi-Dirac one. 

When the atomic magnetic moment is proportional to an atomic spin $S$ which is a multiple of the electron spin~$1 \over 2$, the counting of spin states for each atom can be done by sum rules for the spin. In the case of metals where the moment is a non-integer multiple of $1\over 2$ the sum rules for the spin do not apply. An analytical continuation of the Weiss-Brillouin theory has been used to evaluate the entropy for localized magnetic moments \cite{Tishin-2003}. Further refinements for a theory of the entropy associated with the magnetic moment are obtained by considering the space correlation of the spin fluctuations giving rise to spin waves \cite{Yamada-2003}. This contribution has the same origin of the Bloch law, giving a low temperature correction to the temperature dependence of the saturation magnetization.  It is therefore expected to be relevant at low temperatures~\cite{OHandley-2000}.

\subsection{Ferromagnets}

A simple model for a second order transition is now given in terms of the molecular field theory of ferromagnetism \cite{Chikazumi-1964, Morrish-1965, Herpin-1968,  OHandley-2000}.

\subsubsection{Mean field theory of a ferromagnet}

The ferromagnet is characterised by an exchange interaction between spins which gives an energy term that is minimum for parallel magnetic moments. In the mean field model, the interaction is associated with a molecular field $WM$ which has the dimension of a magnetic field and is proportional to the magnetization $M$. The free energy of a ferromagnet is then:

\beq
f_L = - \frac{1}{2} W \mu_0 M^2 - Ts_M
\label{EQ:fL_fer}
\eeq

\noindent where the first term is the exchange energy, $W$ is the Weiss molecular field coefficient and $s_M$ is the entropy associated with the magnetic moments. By using the expression previously derived for the paramagnet, Eq.(\ref{EQ:s_par}) for the entropy $s_M$, and by introducing the normalised magnetic field $h=H/H_0$, where $H_0 = WM_0$, and the normalized temperature $t=T/T_c$, where $T_c$ is the Curie temperature given by

\beq
T_c = a_J \frac{\mu_0gj\mu_BWM_0}{k_B}\, ,
\label{EQ:Tc_fer}
\eeq

\noindent we find the Weiss equation for ferromagnetism:

\beq
h = - m + t a_J \mathcal{M}_{J}^{-1}(m)\, .
\label{EQ:h_fer}
\eeq

\noindent The stability of the PM and the FM solutions is determined by the condition $\partial h /\partial m > 0$.  By expanding the inverse of the Brillouin function (Eq.(\ref{EQ:Mj_par})) as a power series:

\beq
a_J \mathcal{M}_{J}^{-1}(m) =  m + b_J m^3 + \mathcal{O}(m^5) \, ,
\label{EQ:Mj-1_fer}
\eeq

\noindent we have from Eq.(\ref{EQ:h_fer}) that the paramagnetic state $m=0$ is stable for $t<1$ while the ferromagnetic state $m>0$ is stable for $t<1$. The normalized free energy $\hat{f}_L = f_L /(\mu_0H_0M_0)$ is:

\beq
\hat{f}_L = \hat{f}_0 - \frac{1}{2} m^2 + a_J t s_J \left( m \right)
\label{EQ:fL_norm_fer}
\eeq

\noindent and the first terms of the power expansion in $m$ are

\beq
\hat{f}_L = \hat{f}_0 - \frac{1}{2} m^2 + t \left( \frac{1}{2}m^2+\frac{b_J}{4}m^4 + \mathcal{O}(m^6) \right) \,.
\label{EQ:fL_norm_exp_fer}
\eeq

\begin{figure}[htb]
\narrowtext 
\centering
\includegraphics[width=17cm]{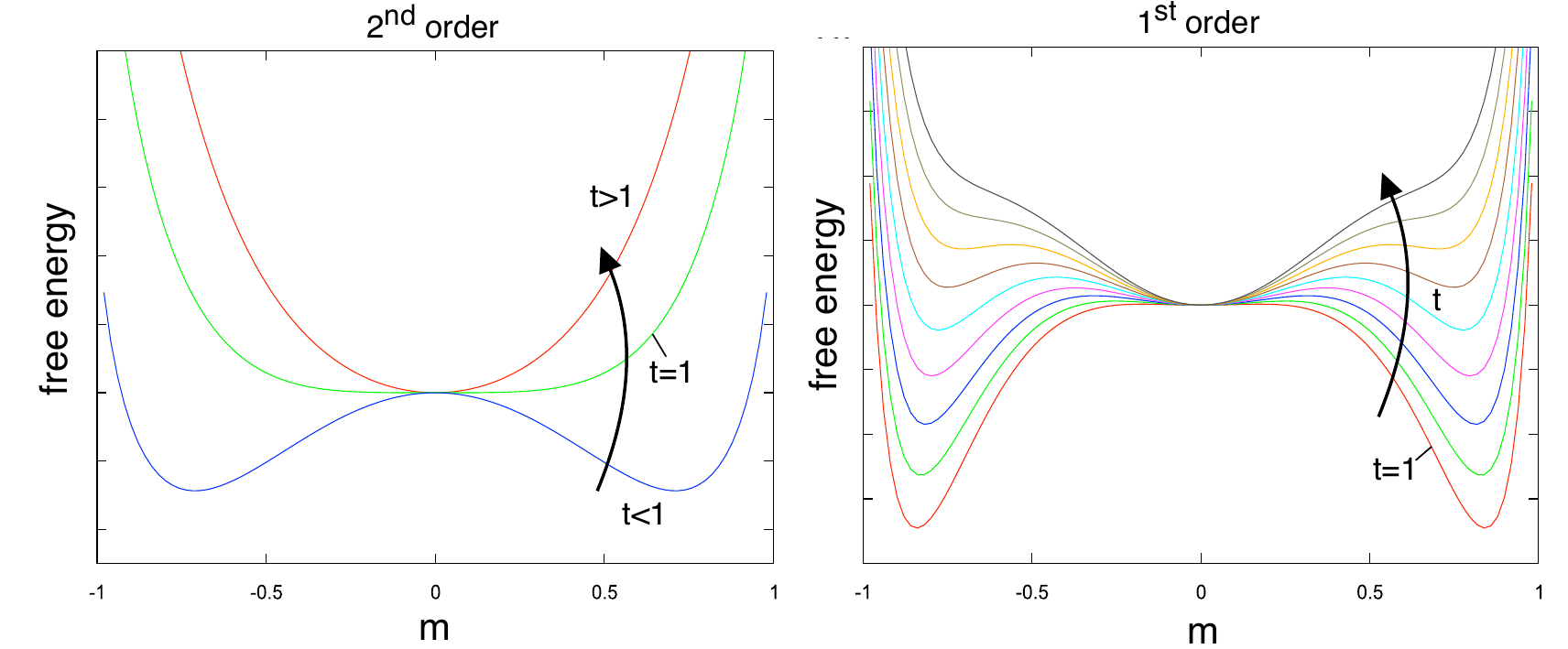}
\caption{Left: Free energy for a second order transition. Right: Free energy for a first order transition} \label{FIG:Free}
\end{figure}

\noindent Fig.\ref{FIG:Free} (left) shows the free energy of Eq.(\ref{EQ:fL_norm_fer}) for different values of $t$. In the example $J=1/2$ for which $\mathcal{M}_{1/2}^{-1}(m) = \tanh^{-1}(m)$ and $s_{1/2}(m)= \ln 2 - ({1}/{2})(1+m) \ln(1+m) - ({1}/{2})(1-m) \ln(1-m)$. 

\subsubsection{Magnetocaloric effect around the second order critical point}

\begin{figure}[htb]
\narrowtext 
\centering
\includegraphics[width=17cm]{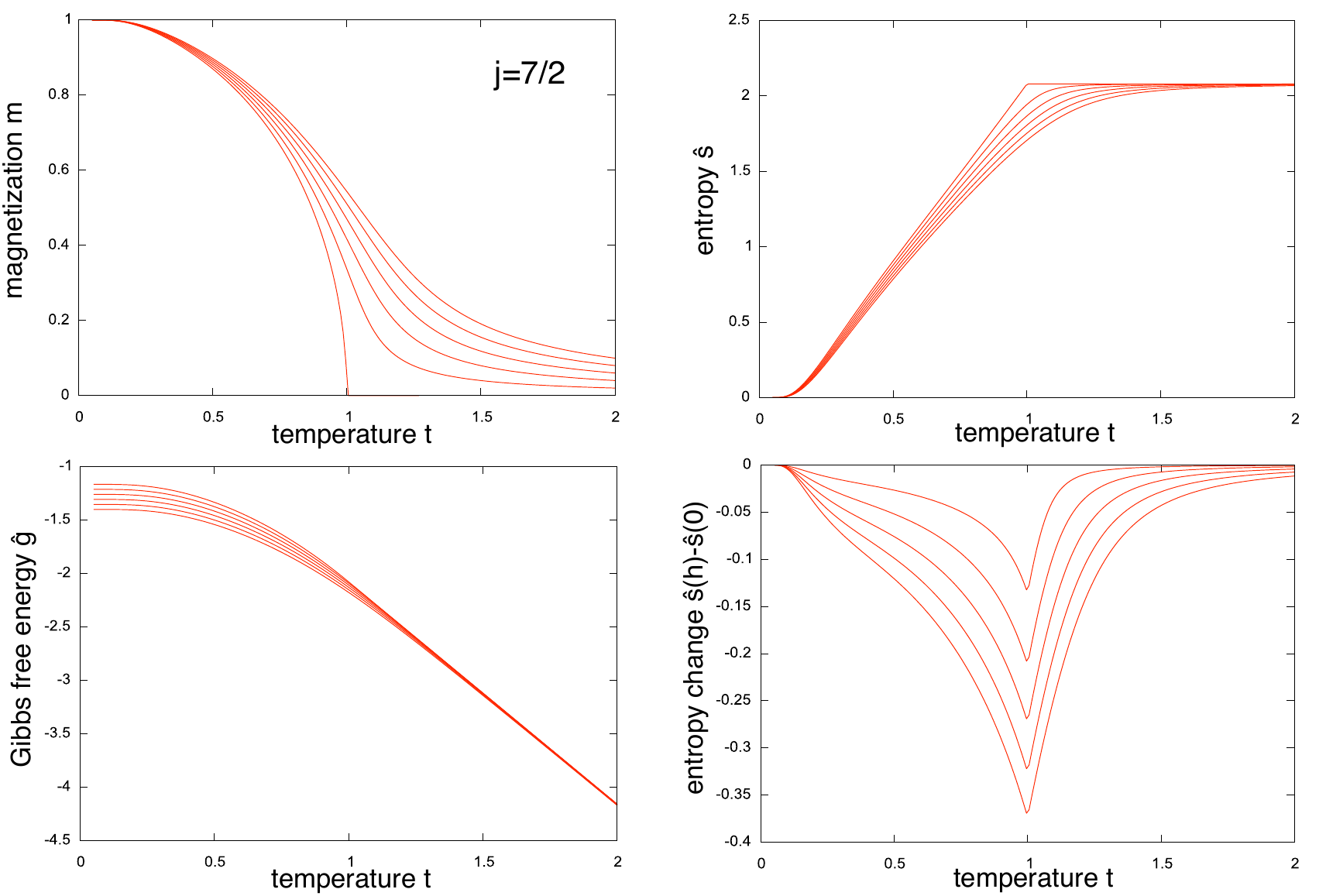}
\caption{Normalized magnetization $m$, entropy $\hat{s}$, Gibbs free energy $\hat{g}=\hat{f}-hm$ and entropy change $\hat{s}(h)-\hat{s}(0)$ for the mean field theory of ferromagnetism with $J={7\over 2}$. The lines correspond to the magnetic field $h$ going from $h = 0$ to $h = 0.05$ in steps of 0.01.} \label{FIG:4_B}
\end{figure}

Once Equation (\ref{EQ:h_fer}) is solved (giving the value of $m$), the entropy can be computed by Eq.(\ref{EQ:s_par}) with the argument $x$ given by $x = \mathcal{M}_{J}^{-1}(m)$. Fig.~\ref{FIG:4_B} shows: the reduced magnetisation $m$ given by the numerical solution of Eq.(\ref{EQ:h_fer}), the reduced entropy $\hat{s} = s_J$ of Eq.(\ref{EQ:sj_par}), the normalized Gibbs free energy $\hat{g}_L=\hat{f}_L-hm$, and the entropy change $\hat{s}(h)-\hat{s}(0)$ for $J=7/2$ where the different lines correspond to the magnetic field $h$ going from $h = 0$ to $h = 0.05$ in steps of 0.01. The magnetic field induced entropy change has a maximum at the Curie temperature, $t=1$. The mean field theory allows us also to derive approximate expressions for the magnetic field induced entropy change $\hat{s}(h)-\hat{s}(0)$ around the Curie temperature $t=1$ and for small $m$. In the paramagnetic state with $t>1$, from Eq.(\ref{EQ:h_fer}), to first order in small $m$ one finds the Curie-Weiss law of the magnetization:

\beq
m \simeq \frac{h}{t-1} \, .
\eeq

\noindent For small $m$ the entropy of Eq.(\ref{EQ:sj_par}) is proportional to $m^2$ and the entropy change has a quadratic dependence on magnetic field:

\beq
\hat{s}(h) - \hat{s}(0) \simeq - \frac{1}{2a_J} \left(\frac{h}{1-t}\right)^{2} \, .
\eeq

\noindent We can see that as $t \rightarrow 1$ entropy change increases. In the ferromagnetic state, $t<1$ and there is a spontaneous magnetization for $h=0$. For small $m$ we have

\beq
m \simeq \left(\frac{1-t}{b_J} \right)^{1/2} \, ,
\eeq

\noindent and the entropy change varies linearly with the field:

\beq
\hat{s}(h) - \hat{s}(0) \simeq -\frac{1}{2a_J}\frac{h}{\sqrt{b_J (1-t)}}
\eeq

\noindent and increases for $t\rightarrow1$. At the Curie point $t=1$ from Eq.(\ref{EQ:h_fer}) we find that the mean field value of the so-called critical exponent with respect to field:

\beq
m \simeq \left(\frac{h}{b_J}\right)^{1/3} \,.
\eeq

\noindent The entropy change is maxised at $t=1$ and varies as the $2/3$ power of the field:

\beq
\hat{s}(h) - \hat{s}(0) \simeq - \frac{1}{2a_J} \left(\frac{h}{b_J}\right)^{2/3} \, .
\eeq

The mean field theory of ferromagnetism presented here can be applied to describe the magnetocaloric effect around the Curie temperature \cite{Tishin-1990}. The MCE has been studied and described with success in ferromagnetic alloys containing rare earth elements \cite{Tishin-2003} and the agreement of the theory with experiments can be further improved by taking into account the crystal field of the surrounding atoms \cite{DeOliveira-2010}. The magnetic field dependences of the entropy change found by the mean field theory around the second order phase transition corresponds well to the exponents found in amorphous alloys \cite{Franco-2006}. Refs.\cite{Kuzmin-2008, Kuzmin-2009, Lyubina-2011} have discussed and extended the entropy change around the Curie temperature in relation to the mean field laws. A more refined approach is obtained by using the theory of critical phenomena around the second order transition. The experimental $\Delta s(H,T)$ values follow the scaling laws of critical phenomena \cite{Franco-2006, Franco-2007} very well in the case of second order Curie transitions.

\section{First order magnetic transitions}

A phase transition is classified as first order when the order parameter (i.e. the magnetization) changes discontinuously. While most magnetic materials have a second order transition at the Curie point, most of the recently developed magnetocaloric materials have discontinuous transitions. Relevant examples include Gd-Si-Ge \cite{Pecharsky-1997, Pecharsky-2003, Zhuo-2003}, Mn-As \cite{DeBlois-1963, Zieba-1982, Pytlik-1985, Rungger-2006, Tocado-2007}, Fe-Rh \cite{Annaorazov-1992, Annaorazov-1996, Manekar-2008}, Mn-Fe-P-As \cite{Zach-1990, Tegus-2002}, Co-Mn-Si \cite{Sandeman-2006}, La-Fe-Si \cite{Fujieda-2004, Gutfleisch-2005, Katter-2008, Moore-2009, Bjork-2010}, mangnanites \cite{Ulyanov-2007, Novak-1999} and the Heusler alloys Ni-Mn-X \cite{Krenke-2005, Krenke-2005b, Pasquale-2007, Sasso-2008, Khovaylo-2010}. First order magnetic phase transitions are the consequence of the coupling between the magnetic moments and the exchange interaction with the electronic and structural degrees-of-freedom. 

\subsection{Coupling between magnetism and structure}

\subsubsection{Exchange energy}

\emph{Indirect exchange}. Much of the recent interest in the MCE in first order magnetic phase transitions has been generated by the discovery of the high values of the magnetic field induced entropy change $\Delta s(H)$ in Gd$_5$Si$_2$Ge$_2$ \cite{Pecharsky-1997}. This effect has been called giant magnetocaloric effect (GMCE) for being larger than the standard material Gd \cite{Dankov-1998}. In Gd$_5$Si$_2$Ge$_2$ the magnetic moment is due to the 4$f$ electrons of the Gd ions which are ferromagnetically coupled by an indirect exchange \cite{Paudyal-2006}. The material has a first order phase transition around 270~K between a low temperature orthorhombic Gd$_5$Si$_4$-type structure and a high temperature monoclinic Gd$_5$Si$_2$Ge$_2$-type structure~\cite{Choe-2000}. The phases have different magnetic properties and the transition can be driven either by the temperature and the magnetic field \cite{Giguere-1999}. Each phase appears to have a different Curie temperature \cite{Magen-2005, Hadimani-2008}. This behavior is confirmed by the fact that  he magneto-structural transition is accompanied by the braking of covalent bonds between Si and Ge, causing a decrease of the exchange interaction between Gd moments. At the transition temperature the variation of the exchange is sufficient to destroy the ferromagnetic order, giving rise to a first order transition \cite{Paudyal-2006} and to a sudden change of the spin entropy. However, it was demonstrated that the change of spin entropy is accompanied by a change of structural entropy, i.e. the monoclinic phase is at higher entropy with respect to the orthorhombic one, giving a structural enhancement of the $\Delta s$ \cite{Magen-2005, Pecharsky-2005, Pecharsky-2009}.

\emph{Volume dependence of direct exchange}. Magnetic first order phase transitions due to the coupling between magnetism and structure were originally discussed to understand  MnAs~\cite{Kittel-1960, Bean-1962}. The magnetic moments of Mn, due to the 3$d$ electrons, would have a natural tendency to align antiferromagnetically, because of a negative exchange integral. However, the exchange is found to increase with the interatomic distance between Mn and, in alloys where the Mn atoms are found at large interatomic distances, it may become positive, giving rise to ferromagnetic order (Ref.~\cite{Coey-2009} p.395). MnAs has a first order phase transition around 312~K between a low temperature hexagonal NiAs-type structure and a high temperature orthorhombic MnP-type structure. The low temperature phase is ferromagnetic while the high temperature phase is paramagnetic. Bean and Rodbell were able to explain the first order phase transition by considering that the ferromagnet exchange depends explicitly on the specific volume~\cite{Bean-1962}. This assumption is reasonable when the change in the unit cell at the transition is reflected in a global volume change. The conclusion is that the low temperature high volume FM phase may collapse into a high temperature low volume PM phase and the order of the transition is governed by the dimensionless parameter $\eta$ which depends on magnetic and structural parameters such as the isothermal compressibility of the lattice $\kappa_T$. When $\eta$ is larger than a critical value $\eta_c$ the transition becomes of first order. The Bean and Rodbell model is particularly interesting as it involves the contribution of the structure to the total entropy. The idea of the model has been developed further in Refs.~\cite{Zach-1998, VonRanke-2004, DeOliveira-2004, VonRanke-2004b, VonRanke-2004c, DeOliveira-2005, VonRanke-2005, Spichkin-2005, Tegus-2005, Balli-2007, Zou-2008, Carvalho-2009, DeOliveira-2010, Basso-2011}.

\emph{Electronic energy} Electronic energy is particularly important for itinerant electrons in metals. In magnetic metals, as for example in transition metals, the 3$d$ electrons contributing to the atomic magnetic moment are not localized at an atomic site. To satisfy the Pauli Exclusion principle, the electronic states of the collective wavefunctions must populate the energy bands of the crystal rather than the atomic levels. In ferromagnetic metals, the energy bands are split into spin-up and spin-down sub-bands and, because some of the electrons are constrained to be spin-polarized as they contribute to the magnetic moment, the spin-up and spin-down energy sub-bands are asymmetrically filled. The asymmetric filling has an energy cost, that has to be balanced by the energy gain due to the ferromagnetic exchange interaction. In a power expansion of the electron energy as a function of the reduced magnetization $m$ one finds the $m^2$ as the first term. This term is inversely proportional to the density of states at the Fermi level $\mathfrak{n}(\epsilon_F)$. 

The result of the energy balance with the exchange energy (which is also proportional to $m^2$, but with the minus sign), is known as the Stoner criterion.  It says that ferromagnetism due to itinerant electrons can exist only if exchange energy dominates over the electronic energy.  By considering a higher order in the power expansion, $m^4$, Wohlfarth and Rodes~\cite{Wohlfarth-1962} demonstrated that if the Stoner criterion is not verified (i.e. if the system is PM), but the coefficient of the $m^4$ order is strong and negative, the system may exhibit a stable coexistence of both PM and FM states at the same temperature. This condition corresponds to the itinerant electron metamagnetic (IEM) transition because a magnetic field may induce a first order phase transition from PM to FM state \cite{Wohlfarth-1962, Shimizu-1982, Levitin-1988, Yamada-2003}. 

The condition of Wohlfarth and Rodes is realised when the density of states is small for the equal filling of spin-up and spin down (PM state) and high for the asymmetric filling of the spin-up band with respect to the spin-down band (FM state). The first order phase transition of La(Fe$_{1-x}$Si$_{x}$)$_{13}$ alloy and its hydrogenation have been explained by invoking this mechanism \cite{Fujita-2003, Fujita-2003b, Kuzmin-2007, Fujieda-2009}. Further contributions to the electronic energy are also expected by the fact that the density of states may depend on the interatomic distance. In particular, close atomic arrangements give rise to wide energy bands characterized by low densities of states. This effect would give rise to an energy contribution in which the effective exchange would depend on volume in a way similar to the magnetovolume argument discussed before in the specific case of Mn \cite{Jia-2006}.

\subsubsection{Structural and electronic free energy}

In solids, the main contributions to the free energy, other than magnetic, are elastic, phononic and electronic bands. The sum of these three terms give rise to the equation of state of a solid for its lattice and electronic parts \cite{Wilson-1953, Wannier-1966, Callen-1985, Landau-1986}. It is worth considering them in detail to have an approximate description of the related free energy terms.

\emph{Elastic energy}. The elastic energy term $f_{ela}$ represents the potential energy related to interatomic forces between the atoms in the lattice and depends on the strain tensor. To a first approximation one may consider isotropic effects and take the elastic energy to be a function of the specific volume change $\omega$. By using a power expansion we have

\beq
f_{ela}(\omega) = \frac{v_0}{2\kappa_0}\omega^2 + \mathcal{O}(\omega^3) \, ,
\label{EQ:f_ela_def}
\eeq

\noindent where $\omega=(v-v_0)/v_0$ is related to the specific volume $v$ and to the specific volume in absence of any pressure $v_0$. 

\emph{Phonon energy}. The phonon term of the structural free energy is due to thermal vibration modes of the atoms in the lattice. In a classical approach, because of the law of the equipartition of the energy, one would have $k_BT$ contribution for each degree-of-freedom of the atom (i.e. 3 for an atom in a solid). In a quantum approach one has to consider the atomic masses as quantum harmonic oscillators and take the spectrum of the vibration modes. A good approximation is given by the Debye model in which the phonon spectrum is taken as isotropic in the wave-vector space with a maximum frequency defined as the Debye frequency $\nu_D$. In the Debye model the free energy of the phonons (\cite{Wannier-1966} p.275) is

\beq
f_D = f(0)+ 3nk_B T \left[ \ln\left[ 1-\exp\left(-y\right)\right]-\frac{1}{3}\mathcal{D}\left(y\right)\right]
\label{EQ:f_D_def}
\eeq

\noindent with

\beq
y = \frac{T_D}{T} \, ,
\eeq

\noindent where $T_D$ is the Debye temperature related to the Debye frequency by $h \nu_D  = k_B T_D$ and $\mathcal{D}(y)$ is the Debye function:

\beq
\mathcal{D}(y) = \frac{3}{y^3}\int_{0}^{y}\frac{x^3}{\exp(x)-1}dx \, .
\eeq

\noindent The Debye model gives a very good description of the specific heat of solids at constant volume, $c_v$
 
\beq
c_v = T\left.\frac{\partial s}{\partial T}\right|_v
\eeq

\noindent where $s = - {\partial f_D}/{\partial T}$ is the entropy.  We thus obtain

\beq
c_v = 3nk_B \mathcal{C}(y)
\label{EQ:f_cv_D}
\eeq

\noindent where

\beq
\mathcal{C}(y) =4 \mathcal{D}\left(y\right) - \frac{3y}{\exp\left(y\right)-1}
\eeq

\noindent is a function that for $T>T_D$ gives $c_v \simeq 3nk_B$, which is the law of Dulong and Petit. The Debye temperature $T_D$ is the only parameter in Eq.(\ref{EQ:f_cv_D}) and is a characteristic of the solid. To describe the thermal expansion in the context of the Debye theory one has to introduce the presence of anharmonic effects of the atomic potential \cite{Dugdale-1953}. When atoms change their interatomic distances, the non linearities of the potential give rise to slight changes of the phonon vibration frequencies. In the quasi-harmonic approximation one still considers harmonic waves, but allows the frequencies to change with the volume $v$. In the Debye model this is introduced through the Gr\"uneisen parameter, $\gamma$:

\beq
\gamma = - \frac{\partial \ln \nu_D}{\partial \ln v}
\eeq

\noindent which expresses the volume dependence of the Debye frequency $\nu_D$. From the definition of the Gr\"uneisen parameter the Debye temperature $T_D(\omega)$ is found to be dependent of the reduced volume $\omega$, introducing a volume dependence in the $f_D$ term.

\emph{Electronic band energy}. In metals, due to the Fermi-Dirac statistics, the contribution of the fluctuations of the electronic states in energy bands is limited to an energy region of amplitude $k_BT$ around the Fermi level. The result of this statistics is found in solid state textbooks \cite{Wilson-1953} and the leading order of a power expansion as a function of the temperature gives the term: 

\beq
f_{ele} = - \frac{\pi^2}{6}n(k_BT)^{2} \mathfrak{n}(\epsilon_F)
\eeq

\noindent where $\mathfrak{n}(\epsilon_F)$ is the density of states of the unsplit band at the Fermi level and the integral of the density of states $\mathfrak{n}(\epsilon)$ up to the Fermi level gives the number of valence electrons per atom. The contribution of such states to the entropy is 

\beq
s_{ele} = \frac{\pi^2}{3}nk_B^2 T \mathfrak{n}(\epsilon_F) \, .
\eeq

\noindent This contribution is linear in $T$ and is often much smaller than the other contributions to the entropy. It is normally relevant to the specific heat only at low temperatures.  There can be exceptions at room temperature, however, when the change of electronic density of states is large during, for example an IEM transition~\cite{Kreiner-1998, Barcza-2013}.

\emph{State equation of a solid}. By using the above approximations to describe an isotropic solid we arrive at an expression for the free energy which is a function of $\omega$ and $T$:

\beq
f_S(\omega,T) = f_{ela}(\omega) + f_D(\omega,T) + f_{ele}(T)\, .
\eeq

\noindent The corresponding equations of state are given by applying Eqs.(\ref{EQ:pf_def}) and (\ref{EQ:sf_def}). By considering the behavior around $\omega=0$ and $T=T_0$ one has the linear state equations:

\beq
p = - \frac{1}{\kappa_T} \omega + \frac{\alpha_p}{\kappa_T} (T-T_0)
\label{EQ:Somega_state}
\eeq
and
\beq
s_S  = s_{S_0} +  \frac{v_0\alpha_p}{\kappa_T} \omega + b_v (T-T_0) \, ,
\label{EQ:Ss_S_state}
\eeq

\noindent which satisfy the Maxwell relation $v_0\partial \omega/\partial T = - \partial s/\partial p$. The parameter $b_v = ds_S/dT|_v$ is the specific entropy capacity at constant volume for the solid related to the specific heat at constant volume by $b_v=c_v/T_0$. We take the elastic term as the first term of the power expansion and a linear expansion of the volume dependence of the Debye temperature $T_D(\omega) = T_{D_0} (1-\gamma\omega)$, defining $y_0 = T_{D_0}/T_0$. We can then derive the values of the parameters appearing in the equations of state (\ref{EQ:Somega_state}) and (\ref{EQ:Ss_S_state}) as a function of the parameters of the elastic, Debye and electronic free energies. The inverse of the isothermal compressibility is 

\beq
\frac{1}{\kappa_T} = \frac{\partial}{\partial \omega} \left( \frac{1}{v_0}\frac{\partial f_S}{\partial \omega} \right)
\eeq

\noindent and yields

\beq
\frac{1}{\kappa_T} = \frac{1}{\kappa_0} - \frac{c_v(y_0)T_0 \gamma^2}{v_0} \, .
\eeq

\noindent The thermal expansion is obtained by:

\beq
\alpha_p = \frac{\kappa_T}{v_0}\frac{\partial}{\partial \omega}\left( - \frac{\partial f_S}{\partial T}\right) 
\eeq

\noindent and is

\beq
\alpha_p = \frac{\kappa_T \gamma c_v(y_0)}{v_0} \, .
\eeq

\noindent The specific entropy capacity at constant volume is given by $b_v=c_v/T_0$ where 

\beq
c_v = 3nk_B \mathcal{C}(y_0) + \frac{\pi^2}{3}nk_B^2 T_0 \mathfrak{n}(\epsilon_F)
\eeq

\noindent is the specific heat at constant volume. By taking the three parameters $\kappa_T$ $\alpha_p$ and $b_v$ as constants, the free energy $f_S(\omega, T)$ can be expressed as a power expansion around the values $\omega=0$ and $T=T_0$ \cite{Landau-1986}:

\beq
f_S(\omega,T) = f_{S_0}(0,T_0)+\frac{v_0}{\kappa_T}\frac{\omega^2}{2} - \left[ \frac{\alpha_p v_0}{\kappa_T}\omega + s_{0} \right] (T-T_0) - b_v\frac{1}{2} (T-T_0)^2 \, .
\label{EQ:f_e}
\eeq

\noindent The linear state equations for the reduced volume $\omega$ and the entropy of the structural part $s_S$, valid around $p=0$ and $T=T_0$, are:

\begin{eqnarray}
\label{EQ:omega_state}
\omega &=& - \kappa_T p + \alpha_p (T-T_0)\\
s_S - s_{S_0} &=& - v_0 \alpha_p p + b_p (T-T_0)
\label{EQ:s_S_state}
\end{eqnarray}

\noindent where $s_{S_0}$ is a reference entropy value (at $T=T_0$ and $p=0$) and $b_p$ is related to $b_v$ by the expression $b_v = b_p - \alpha_p^2v_0/\kappa_T$ and is related to the specific heat at constant pressure by $b_p=c_p/T_0$. The corresponding Gibbs free energy the structural lattice is finally

\beq
g_S(p,T)=g_{S}(0,T_0)-\frac{1}{2}v_0\kappa_Tp^2+(v_0p\alpha_p-s_0)(T-T_0)-\frac{1}{2}b_p(T-T_0)^2
\eeq

\subsection{First order transition due to magneto-elastic coupling}

The paradigm for a first order magnetic transition is arguably the Bean and Rodbell model of magneto-elastic coupling \cite{Bean-1962}. The basic idea of the model is to describe a ferromagnet in which the the interatomic distance influences the exchange interaction. If the change of interatomic distance is reflected in a global volume change the ferromagnetic exchange depends explicitly on the volume and one has a coupling between the elastic and magnetic parts of the free energy. The first order nature of the transition is revealed by the minimization of the total free energy due to the sum of these free energy terms. The Bean and Rodbell model is a paradigm example for the magnetocaloric effect because it shows how the entropy of the crystal lattice may be involved in the magnetic field-induced total entropy change.

\subsubsection{The Bean-Rodbell model}

The specific Landau free energy $f_L(M,\omega,T)$ is:

\beq
f_L = - \frac{1}{2} W(\omega) \mu_0 M^2 - Ts_M(M) + f_S(\omega, T)
\label{EQ:fL_BR_def}
\eeq

\noindent where the first two terms on the right hand side are the free energy of the ferromagnet, Eq.(\ref{EQ:fL_fer}), and $f_S(\omega, T)$ is the free energy describing the structural lattice. The molecular field coefficient $W$ is assumed to depend linearly on the reduced volume as $W(\omega) = W_0(1+\beta\omega)$ where $\beta$ is a dimensionless coefficient. The basic result of the Bean and Rodbell model can be obtained by the approximated $f_S(\omega, T)$ of Eq.(\ref{EQ:f_e}), giving linear equations of state for the structural part of the system. The state equations for the magneto-elastically coupled magnetic material are given by imposing both Eq.(\ref{EQ:pf_def})

\beq
\frac{1}{v_0}\frac{\partial f_L}{\partial \omega} = - p
\eeq

\noindent and Eq.(\ref{EQ:Hf_def})

\beq
\frac{\partial f_L}{\partial M} = \mu_0 H
\eeq

\noindent By imposing the first condition we obtain the equilibrium value of $\omega$:

\beq
\omega = - \kappa_T \left( p - \frac{\eta}{3\beta\kappa_T}m^2 \right)  + \alpha_p (T-T_0) \, ,
\label{EQ:omega_sol}
\eeq

\noindent where we have introduced the dimensionless parameter $\eta$ of Bean and Rodbell~\cite{Bean-1962}:

\beq
\eta = \frac{3}{2} \frac{\beta^2 \kappa_T \mu_0 M_0^2W_0}{v_0} \, .
\label{EQ:eta_def}
\eeq

\noindent By comparing Eq.(\ref{EQ:omega_sol}) with Eq.(\ref{EQ:omega_state}) we see that the volume dependence of the ferromagnetic exchange gives rise to an exchange magnetostriction term which appears as an equivalent pressure $p_W = - {\eta}\,m^2/({3\beta\kappa_T})$.  This depends on the square of the magnetization, $m^2$. By imposing the second condition we obtain the equation 

\beq
-N_W(\omega) \mu_0 M - T \frac{\partial s_M}{\partial M} = \mu_0 H \, .
\eeq

\noindent By substituting $\omega$ from Eq.(\ref{EQ:omega_sol}) and dividing all terms by $\mu_0H_0 = \mu_0M_0W_0$ we have 

\beq
h = - \left[ 1 + \beta ( \alpha_p (T-T_0) -  \kappa_T p ) \right]  m - \frac{1}{3} \eta m^3 - t \frac{a_J}{nk_B} \frac{\partial s_M}{\partial m}
\label{EQ:h}
\eeq

\noindent where $t=T/T_{c_0}$ and $T_{c_0}$ is given by Eq.(\ref{EQ:Tc_fer}). The temperature $T_0$ of Eqs.~(\ref{EQ:omega_state}) and (\ref{EQ:s_S_state}) is arbitrary.  Then, by taking $T_0=T_{c_0}$, we may write the linear $m$ term as $- \left[ 1 + \zeta (t-1) -  \pi \, \right]  m$ where we define the dimensionless pressure $\pi=\beta \kappa_T p$ and the dimensionless parameter $\zeta$ (zeta),

\beq
\zeta = \alpha_p \, \beta \, T_{c_0}
\label{EQ:zeta}
\eeq

\noindent which takes into account the role of the thermal expansion of the lattice. The normalized Landau free energy $\hat{f}_L(m,t)$ as a function of $m$ is obtained by the integral of Eq.~(\ref{EQ:h}):

\beq
\hat{f}_L = - \frac{1}{2}  \left( \left[ 1 + \zeta (t-1) -  \pi \, \right]  m^2 + \frac{1}{6} \eta m^4 \right) - t \frac{a_J} {nk_B} s_M(m) \, .
\label{EQ:fL_norm}
\eeq

\noindent The magnetization $m$ is given by the solution of Eq.(\ref{EQ:h}):

\beq
h = - \left[ 1 + \zeta (t-1) -  \pi \, \right]  m - \frac{1}{3} \eta m^3 + t \,a_J\,\mathcal{M}_{J}^{-1}(m) \, .
\label{EQ:h_eq_1}
\eeq

\noindent The number of possible stable solutions of Eq.(\ref{EQ:h_eq_1}) is evaluated by taking the power expansion. One obtains: 

\beq
h = \left[ (t-1)( 1- \zeta) +\pi \right] m + \left( t \, b_J - \frac{\eta}{3}  \right) m^3 + t \mathcal{O}(m^5) \, .
\label{EQ:h_exp}
\eeq

\noindent By defining 

\beq
t_{P}=1-\frac{\pi}{1-\zeta}
\eeq

\noindent we have that, for $h=0$ the PM state with $m=0$ is always a solution. But the PM state is an energy minimum only for $t>t_P$, while for $t<t_P$ there is always one stable solution with $m>0$, i.e. a FM state \cite{Basso-2011}. The order of the transition is determined by the sign of $(t_P b_J - {\eta}/{3})$. By defining the critical value $\eta_c = 3 b_J t_P$ we have that when the PM solution is marginally stable ($t=t_P$), there is a FM solution if $\eta > \eta_c$.  This means that the PM and FM states may coexist and the transition is first order. The normalized Landau free energy $\hat{f}_L(m,t)$ as a function of $m$, Eq.~(\ref{EQ:fL_norm}), is shown in Fig.~\ref{FIG:Free} for $J={1 \over 2}$, $t_P=1$, $\eta_c=1$ and $\eta=2$ for different values of $t$, showing the coexistence of PM and FM states. If $\eta < \eta_c$ there is no possible coexistence and the transition is instead second order. 


\subsubsection{Magnetocaloric effect around the first order phase transition}

The entropy is given by 
\beq
s = - \left.\frac{\partial f_L}{\partial T}\right|_{m,\omega}
\eeq

\noindent By taking the derivative of Eq.(\ref{EQ:fL_BR_def}) with respect to $T$ and substituting Eq.(\ref{EQ:omega_sol}) we obtain

\beq
s = s_M(m) + s_{W}(m) +s_{S}(p,T)
\eeq

\noindent where $s_M(m)$ is the magnetic entropy of Eq.(\ref{EQ:s_par}), $s_{S}(p,T)$ is the structural lattice entropy of Eq.(\ref{EQ:s_S_state}) and 

\beq
s_{W}(m)= \frac{n k_B}{2 a_J} \zeta m^2
\eeq

\noindent is the \emph{magneto-elastic} entropy, a term of structural lattice origin, induced by the ferromagnetic exchange forces through the magneto-elastic interaction. The magnetic entropy $s_M(m)$ has a maximum at $m=0$ and it decreases to zero for $m=1$. The magneto-elastic entropy depends on the parameter $\zeta$ and is proportional to $m^2$. To analyse the competition between $s_M$ and $s_W$, the two terms that depend on $m$, we introduce the normalized entropy, $\hat{s}(m) = (s_M(m)+s_W(m))/(nk_B)$. The maximum difference is between the entropy at $m=0$ and $m=1$, $\Delta \hat{s}_{max}=\hat{s}(0)-\hat{s}(1)$ and is:

\beq
\Delta \hat{s}_{max} = \ln(2J+1)-\frac{1}{2 a_J}\zeta \, .
\label{EQ:s_sat}
\eeq

\noindent By using the power expansion of Eq.(\ref{EQ:sj_exp_par}) for $s_M$ we obtain 

\beq
\hat{s} = \ln(2J+1) - \frac{1}{2a_J}\left[ \left(1-\zeta\right) m^2 + \frac{b_J}{2}m^4 + \mathcal{O}(m^6) \right]
\label{EQ:s_red}
\eeq

\noindent where we see that the total entropy may be increased or decreased depending on the sign of $\zeta$. When $\eta>\eta_c$ the transition is first order and there is a discontinuous jump of the magnetization $m$. At the transition temperature between the low temperature phase (LT) and the high temperature phase (HT), the entropy $\hat{s}$ increases discontinuously with a jump $\Delta s = s_{HT}- s_{LT}>0$. We therefore have the following cases:

\begin{itemize}

\item for $\zeta<1$ the transition is from LT-FM ($m\neq0$) to HT-PM ($m=0$) and the magnetic entropy change is positive, $\Delta s_M>0$:

\begin{itemize}
\item For $\zeta<0$ the magneto-elastic entropy change is positive, $\Delta s_W>0$, and there is an enhancement of the total entropy change with respect to the magnetic contribution $\Delta s>\Delta s_M$.
\item For $0<\zeta<1$ the magneto-elastic entropy change is negative, $\Delta s_W<0$, and there is reduction of the total entropy change with respect to the the magnetic contribution $\Delta s<\Delta s_M$. 
\item For $\zeta \rightarrow 1$, to order $m^2$ the two contributions oppose one other $\Delta s_W \rightarrow - \Delta s_M$, and $\Delta s\rightarrow 0$
\end{itemize}

\item For $\zeta>1$ the transition is from LT-PM ($m=0$) to HT-FM ($m\neq0$) and the magnetic entropy change is negative $\Delta s_M<0$, but the magneto-elastic entropy change is positive. For $\zeta>\zeta_c$, where $\zeta_J=2a_J\ln(2J+1)$ is the critical value at which the entropy of the $m=0$ and $m=1$ are the same, we have that $\Delta s_W>-\Delta s_M$, and at the transition the total entropy change is lower than the magneto-elastic contribution $\Delta s < \Delta s_W$ \cite{Basso-2011}.
\end{itemize}

\begin{figure}[htb]
\narrowtext 
\centering
\includegraphics[width=18cm]{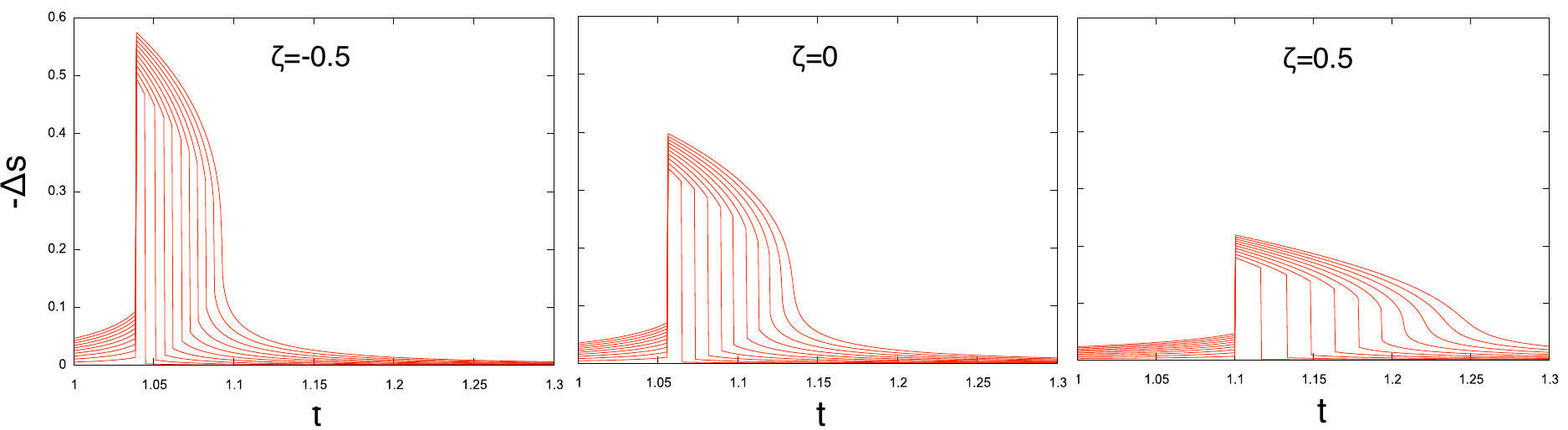}
\caption{Magnetic field induced entropy change $\Delta s(h,t)$ for the Bean Rodbell model computed for $J=1/2$. The lines correspond to magnetic field $h$ from 0 to 0.04 in steps of 0.004.} \label{FIG:entropyB}
\end{figure}

In a first order transition the equilibrium is determined by the Maxwell convention in which the system is allowed to select the minimum with the lowest Gibbs free energy. The Gibbs potential is $g_L=f_L-\mu_0HM+pv_0\omega$. By taking the difference $g_L-g_S$, and dividing by $\mu_0H_0M_0$ we obtain the normalized potential $\hat{g}_L = \hat{f}-hm$, where $\hat{f}$ is given by Eq.(\ref{EQ:fL_norm}). Fig.~\ref{FIG:entropyB} shows the magnetic field induced entropy change computed for $J={1 \over 2}$, under $p=0$ for which $\mathcal{M}_{J}^{-1}(m) = \tanh^{-1}(m)$, $a_J=1$, $b_J=1/3$ and $\eta_c=1$. The values of the parameters are $\eta=2$, $\zeta=-0.5, \,0.0,\, 0.5$. The magnetic field $h$ is in the range $0 < h < 0.04$ in $h$ steps of 0.004. Eq.~(\ref{EQ:h_eq_1}) is solved numerically and the transition is taken at the temperature at which $\hat{g}_{FM}=\hat{g}_{PM}$. Fig.~\ref{FIG:entropyB} shows that the entropy change increases when the contribution from the structure is positive (at negative $\zeta$) and that the the transition temperature $t_t$ dependence on the magnetic field $h$ decreases. This is due to the Clausius-Clapeyron equation which, in normalized form, is $dt_t/dh = \Delta m / \Delta \hat{s}$, where $\Delta m$ is the discontinuous change of the magnetization. At a given value of $\Delta m$ a lower $\Delta \hat{s}$ corresponds to a higher $dt_t/dh$ which may result in a higher adiabatic temperature change.

The coupling between volume and magnetism has been studied by Jia \emph{et al.} \cite{Jia-2006}, who noticed that if at the transition the volume changes discontinuously, then the total entropy change, including the magnetic contribution, is affected. To sustain this idea, in Refs.\cite{Jia-2006, Liu-2006, Duan-2008} the authors have represented the experimental entropy change $\Delta s$ of several magnetocaloric alloys (Gd$_{5}$Si$_{x}$Ge$_{1-x}$, LaFe$_{13-x}$Si$_{x}$ and Ni$_{2.15}$Mn$_{0.85-x}$Cu$_{x}$Ga) as a function of the magnetization changes. By subtracting from this data the theoretical magnetic entropy, they were able to determine the sign and the amplitude of the structural contribution. In Gd$_5$Si$_2$Ge$_2$ and Ni$_{2.15}$Mn$_{0.85-x}$Cu$_{x}$Ga the lattice contribution increases the entropy change at the transition while in LaFe$_{13-x}$Si$_{x}$ and MnAs \cite{Bean-1962}, even if the magnetic entropy change is large, the role of the lattice entropy is that of decreasing the $\Delta s$. This has not necessarily negative implication for the magnetocaloric effect, because, as shown in the example of Fig.\ref{FIG:entropyB} the temperature width of the $\Delta s$ increases. 

\subsection{Magnetocrystalline anisotropy energy}

The description of the anisotropic nature of the MCE has received particular attention in the research literature~\cite{Bennett-1993, Tishin-2003, DeOliveira-2010}. The rotation of the magnetization away from its easy axis has an associated entropy change which is due to the different spin entropy along different directions and is phenomenologically described by the temperature dependence of the anisotropy constants. For example by taking the uniaxial anisotropy to first order

\beq
f_{AN} = K_1(T)\sin^2\theta
\eeq

\noindent where $\theta$ is the angle formed by the magnetization vector with the easy axis. The entropy term associated with the anisotropy will be

\beq
s_{AN} = - \frac{df_{AN}}{dT} =- \frac{dK_1}{dT}\sin^2\theta
\eeq

\noindent and therefore the entropy difference between the $\theta=\pi/2$ and the $\theta=0$ directions will be $\Delta s_{AN} = s_{AN}(\pi/2)-s_{AN}(0) = - {dK_1}/{dT}$. In ferromagnets with an easy axis ($K_1>0$) the entropy change associated with the rotation of the magnetization along the hard direction (hard plane) gives an increase of the entropy if ${dK_1}/{dT} < 0$. 

The underlying physical phenomenon is that the entropy of the spin system is larger if the magnetization is directed along a hard direction. A basic understanding of this phenomenon can be obtained by considering the Callen and Callen law of the anisotropic magnetization~\cite{Callen-1960}. If the total magnetization is constrained along a direction of hard magnetocrystalline anisotropy, the atomic magnetic moments will tend to fan out around the hard axis in order to minimize the total energy. This effect gives rise to an averaging of the magnetocrystalline anisotropy and a decrease of the saturation magnetization value along the hard axis (anisotropic magnetization). If we simply associate spin disorder with spin entropy, we then obtain that the spin entropy will be larger along an hard axis. First principles evaluation of this anisotropic contribution to the entropy requires specific theoretical developments \cite{Kuzmin-1995, Kuzmin-2007b, DeOliveira-2010}. 

The Callen and Callen argument may help us to understand the magnetocaloric effect in spin reorientation transitions in the presence of two magnetic sublattices, as for example in Er$_2$Fe$_{14}$B \cite{Hirosawa-1985, Kou-1995, Pique-1996, Skokov-2009, Basso-2011b} and NdCo$_5$ \cite{Nikitin-2010} and in other alloys \cite{Ohkoshi-1977, Pedziwiatr-2005, Ilyn-2008, Ilyn-2009}. If the moments of the two sublattices are rigidly coupled (ferro or antiferro), then minimization of the total energy will select which sublattice will satisfy its local anisotropy.  At high temperature the system will always be in the state that yields the highest entropy. For both Er$_2$Fe$_{14}$B and NdCo$_5$ it is the RE moments that dominate the entropy contribution at high temperature, probably because they are loosely coupled to each other \cite{Herbst-1991}. The reorientation transition can be discontinuous (from plane to axis) as well as continuous (through an intermediate easy cone) depending on the high order anisotropy constants \cite{Asti-1980, Nikitin-2010, Basso-2011b}.

\section{Hysteresis and modeling}

In the previous sections we have considered equilibrium first order phase transitions by using the Maxwell convention in which the system selects the energy minimum of lowest energy. This is however only an idealized limit situation. As a matter of fact, real systems do not follow either the equilibrium transition or the completely out-of-equilibrium picture given by the global instabilities of the dashed lines of Fig.~\ref{FIG:fL} (center).  Instead they behave in an intermediate way~\cite{Planes-1992}. The first order transition occurs by the spontaneous formation of domains of the new phase within the old phase. The domains will be separated by phases boundaries and the phase transformation may occur by the motion of these boundaries in a phase coexistence state. 

In real systems many internal non-intrinsic contributions play a major role. These effects give rise to: i) a smooth transition between the phases rather than the vertical slope of the equilibrium Maxwell construction, and ii) a hysteresis with smaller amplitude with respect to the jumps of the global instability picture. The formation of the nuclei of the new phase is somehow spread around the Maxwell construction because phase coexistence may contribute to the minimization of space-dependent energy terms such as elastic energy, related to the internal stresses, and magnetic energy, related to internal magnetostatic fields. These effects are therefore related to the presence of structural defects and disorder. The distribution of disorder also gives rise to localized energy barriers for the nucleation and the motion of the phase boundaries which are smaller than the energy barrier separating the two minima of the free energy.

\subsection{Hysteresis and entropy production}

\begin{figure}[htb]
\narrowtext 
\centering
\includegraphics[width=15cm]{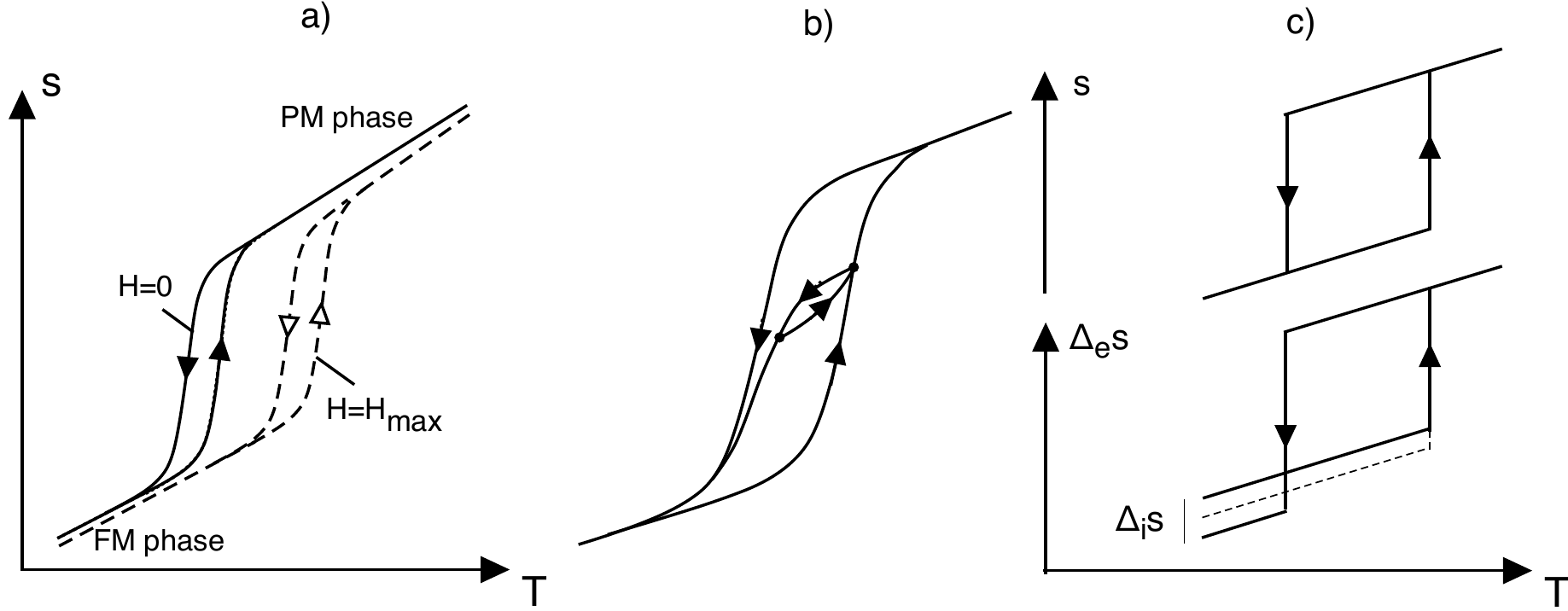}
\caption{a) Entropy as a function of temperature and magnetic field in a first order transition with hysteresis. b) Branching example. c)Top: entropy of a squared hysteresis loop. Bottom: integral of the exchanged entropy in an idealized heating and cooling experiment.} \label{FIG:E1}
\end{figure}

The entropy in the presence of a first order, hysteretic transition is sketched as a function of temperature in Fig.~\ref{FIG:E1}a. The presence of hysteresis has the peculiar effect, making the magnetocaloric properties history-dependent. Both the entropy change and the temperature change depend on the history of the $H$ and $T$ variables in preparing the experimental material sample~\cite{Kuepferling-2008, Basso-2010b}. The presence of reversible and irreversible effects is clearly revealed in the measurement of the specific heat, which is different if it is measured by temperature scanning experiments or by ac experiments~\cite{Morrison-2010}. The reason is that in phase transitions with hysteresis there is a superposition of irreversible and reversible processes. While the scanning experiments catch all processes the ac methods select only the reversible ones. Fig.~\ref{FIG:E1}b shows how the change of the direction of the temperature variation corresponds to tracing of a new entropy-temperature curve and a further reversal produce a minor hysteresis loop \cite{Bertotti-1998}. The direct application of equilibrium relations such as the Maxwell relations to first order phase transitions with hysteresis may then create ambiguous results as discussed widely in the literature \cite{Basso-2005, Basso-2006, Liu-2007, Basso-2008c, Caron-2009, Tocado-2009, Amaral-2009, Balli-2009}. These problems can be avoided by using direct calorimetric methods \cite{Pecharsky-1997b, Plackowski-2002, Marcos-2003, Jeppesen-2008, Miyoshi-2008, Klaasse-2008, Basso-2008, Bratko-2012}. 

A second point worthy of discussion here is the fact that, in an out-of-equilibrium process we also have to deal with the non-conservation of the entropy~\cite{Callen-1985, Bertotti-2006}. For an out-of-equilibrium process the second law of thermodynamics is stated as $\delta s = \delta_e s + \delta_i s$ where $\delta s$, the differential of the entropy state variable, equals the sum of $\delta_e s$ the differential of the entropy exchanged with the surrounding thermal bath, and $\delta_i s$, the differential of the entropy produced internally by irreversible processes. The entropy exchanged with the thermal bath can be estimated from a direct measurement because $T\delta_e s/dt = dq/dt$ is the heat flow with the thermal bath. For the evaluation of the entropy $s$ of a material with hysteresis one should be able to evaluate both the instantaneous entropy production $\delta_i s$ and the exchanged entropy $\delta_e s$. The entropy production $\delta_i s$, is definite positive as a consequence of the second principle of thermodynamics but it can only be measured in a cyclic process. In a closed cycle transformation we have $\oint \delta s = 0$, therefore the entropy produced over one entire loop is $\Delta_i s = \oint \delta_i s = - \oint \delta_e s $. The differential $\delta_i s$ cannot be determined by purely experimental means and a physical theory separating the exchanged and produced entropy is needed in order to compute $s$ from measured heat flux~\cite{Basso-2010}. To have an order of magnitude of the two, we note that the amplitude of $\Delta_i s$ is independently given by the heat dissipated in a $s$ vs. $T$ hysteresis loop which is given by the loop area $\oint s dT$. If we approximate the $s$ vs. $T$ loop as a parallelogram of height $\Delta s$ and width $\Delta T_{hyst}$ (see Fig.~\ref{FIG:E1}c), we have that the entropy production over the entire loop is approximately $\Delta_i s = \Delta s \Delta T_{hyst}/T$ where $T$ is the average temperature of the transition. As the entropy production is definite positive, the measurable integral $\Delta_e s = \int (\delta s - \delta_i s) dt$ will have the shape shown in Fig.~\ref{FIG:E1}c. The entropy produced in the entire loop depends on the ratio $\Delta T_{hyst}/T$ which, for magnetocaloric materials with transitions around room temperature $T \simeq 300$ K and small temperature hysteresis $T_{hyst}<1$ K, is a small contribution that may be disregarded to a first approximation \cite{Sasso-2005, Basso-2006b, Basso-2007b}. 

\subsection{Equivalent driving force}

In order to arrive at a model of the hysteresis in the first order phase transition we first consider a non-equilibrium Gibbs free energy $g_L(M;H,p,T)$ that, in a certain range of its intensive parameter ($p$, $H$, or $T$), is characterized by two distinct energy minima as a function of the magnetization $M$ (taken here as the order parameter). All the other extensive variables, the volume $v$ and the entropy $s$, are related to $M$ as in the example of Section III.A. The two minima of the function $g_L(M;H,p,T)$ correspond to the two stable  phases that we can call the low temperature phase (LT) and the high temperature phase (HT), depending on their relative stability with respect to $T$.  We may then consider the thermodynamics of each of the phases separately. 

We examine the non-equilibrium Gibbs free energy at each minimum, i.e. $g_{LT} = g_L(M_{LT};H,T)$ and $g_{HT} = g_L(M_{HT};H,T)$. In a limited range of $H$ and $T$, the energies $g_{LT}(H,T)$ and $g_{LT}(H,T)$ can be considered as the equilibrium potential. This occurs as soon as there exists an energy barrier separating the two minima. Once we have defined this initial hypothesis, we consider the phase transition between the LT phase and the HT phase driven by either of the intensive variables ($p$, $H$, or $T$). In thermodynamic equilibrium the Maxwell construction would apply and the system would select the state for which the Gibbs free energy is minimum. In presence of two phases LT and HT with different Gibbs potentials $g_{LT}$ and $g_{HT}$ respectively, the sign of the difference $g_{LT}-g_{HT}$ which will decide which of the two phases is globally stable. If $g_{LT}-g_{HT}<0$ the system will be in the LT phase, while if $g_{LT}-g_{HT}>0$ the system will be in the HT one. The free energy difference $g_{LT}-g_{HT}$ takes then the role a driving force of the transformation, encapsulating the action of temperature,  pressure and magnetic field~\cite{Basso-2007, Shamberger-2009}.

\subsection{Preisach-type models}

\begin{figure}[htb]
\centering
\includegraphics[width=18cm]{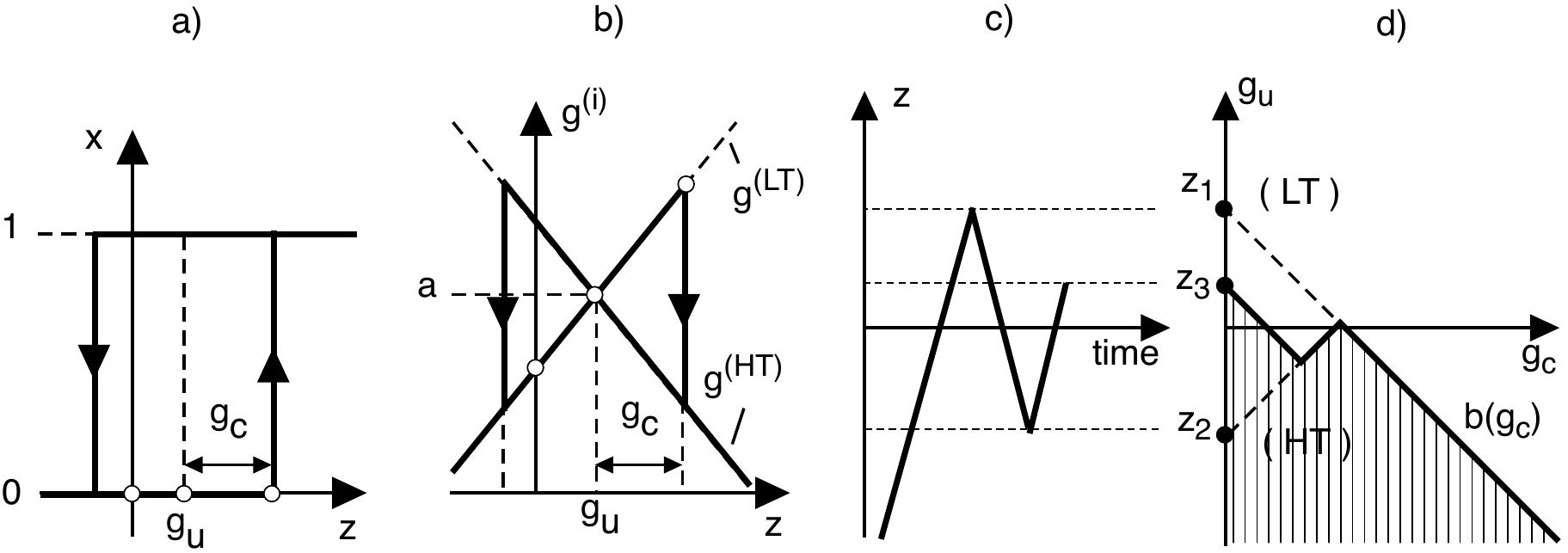}
\caption{a) A bistable unit of the phase transformation of the phase fraction $x$ as a function of effective force $z = (g_{LT} - g_{HT})/2$. $g_{c}$ and $g_{u}$ represent the effect of structural disorder. b) Energy of the bistable unit. c) Temporal hstory of the input $z(H,T)$. d) State line $b(g_c)$ in the Preisach plane $(g_c,g_u)$ representing the state of an ensemble of bistable units.}
\label{FIG:H}
\end{figure}

The presence of disorder gives rise to a complex hysteresis relationship characterized by smoothed, rather than abrupt, properties and the phenomenon of branching at the turning points of the input variable~\cite{Bertotti-1998}. Hysteresis has been studied in detail in particular by using a Preisach-type model in which the output is due to the superposition of many bistable units~\cite{Bertotti-2006}. To describe a first order phase transformation in terms of bistable contributions we consider as a driving force the half difference $z(H,T) = (g_{LT}-g_{HT})/2$~\cite{Basso-2007}. Each unit has switching thresholds at $z = g_u \pm g_c$ where the $+$ sign refers to the switch from $0 \rightarrow 1$ and the $-$ sign to $1 \rightarrow 0$.  The values of $g_{u}$ and $g_c$ are properties of the individual unit (Fig.~\ref{FIG:H}a). The units are distributed according to two parameters: the width $g_c$ and the shift $g_u$. Here we suppose that $g_{u}$ and $g_c$ are independent of the intensive variables, reflecting the effects of structural disorder only. The disorder in the material is reflected in a statistical distribution of the units, $p(g_{c}, g_{u})$. At a given instant of time, the state (0 or 1) of each bistable unit can be represented in the $(g_c,g_u)$ plane and the regions of the plane in the 0 or 1 state are determined by the temporal history of $z(t)$ only. The approach to the out-of-equilibrium phase transformation just described turns out to be perfectly equivalent to the Preisach model of hysteresis. The $p(g_{c}, g_{u})$ distribution is then called the Preisach distribution and all the mathematical results of that model can be applied to the present case. In particular, in the plane $(g_c,g_u)$ the 0 and 1 regions are separated by the borderline function $b(g_{c})$ (Fig.~\ref{FIG:H}d) which is determined by the temporal history of $z(t)$ (Fig.~\ref{FIG:H}c) by the inequality $\left | b(g_{c}) - z(t) \right | \leq g_{c}$ at each time instant. This borderline function fully characterizes the non-equilibrium phase-coexistence state of the material.

\begin{figure}[htb]
\centering
\includegraphics[width=18cm]{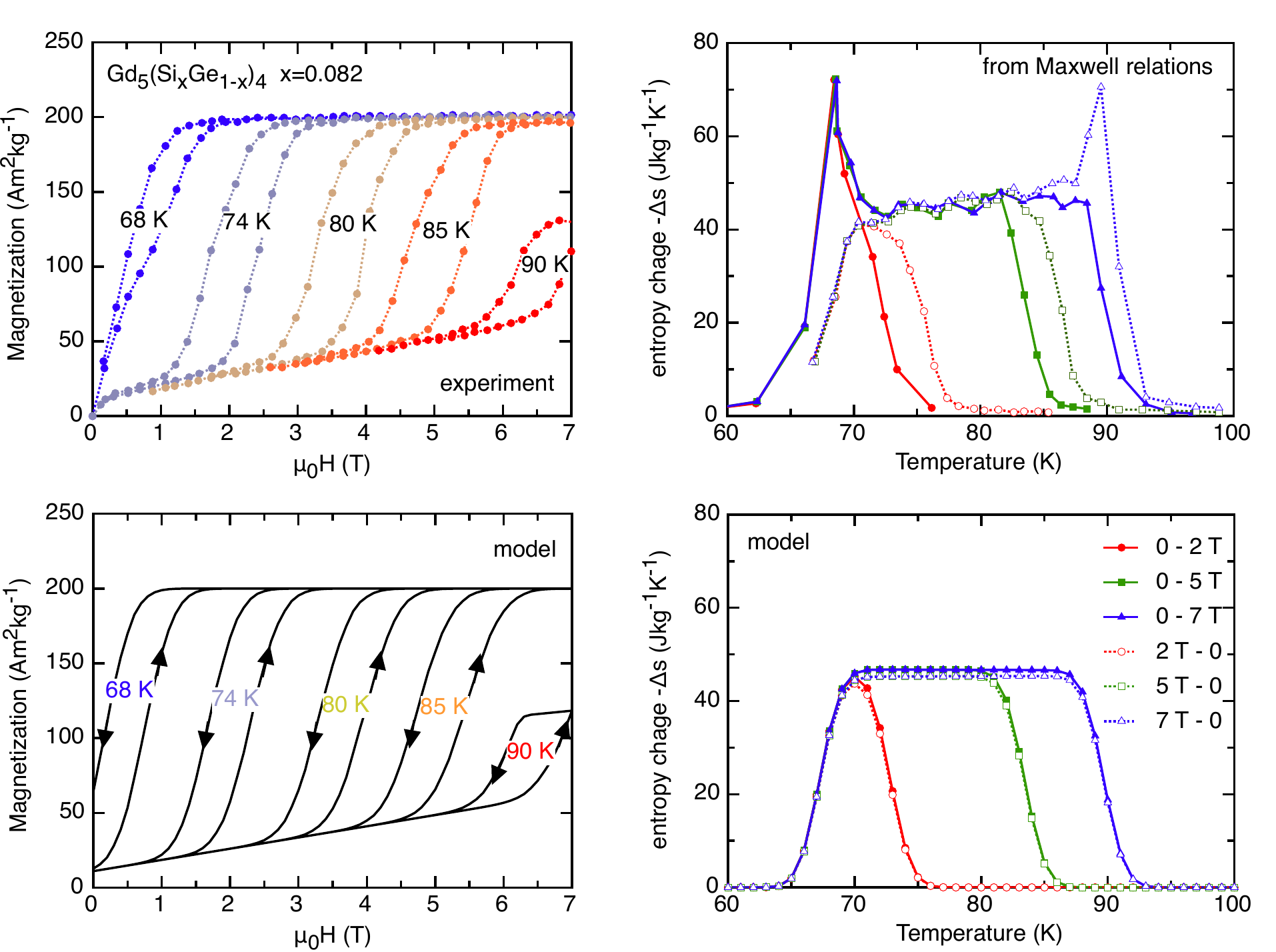}
\caption{A model of Gd$_5$(Si$_x$Ge$_{1-x}$)$_4$ with $x=0.082$. Top left: experimental $M(H,T)$ curves after Ref.~\cite{Zhuo-2003}.  Top right: $\Delta s$ computed from the Maxwell relation using the experimental data, after Ref.~\cite{Zhuo-2003}. Bottom left: model of $M(H,T)$.  Bottom right: prediction of $\Delta s$ from the model fter Ref.\cite{Basso-2006}. The model does not predict the unphysical spikes obtained with Maxwell relations.}
\label{FIG:Maxwell}
\end{figure}

\begin{figure}[htb]
\centering
\includegraphics[width=18cm]{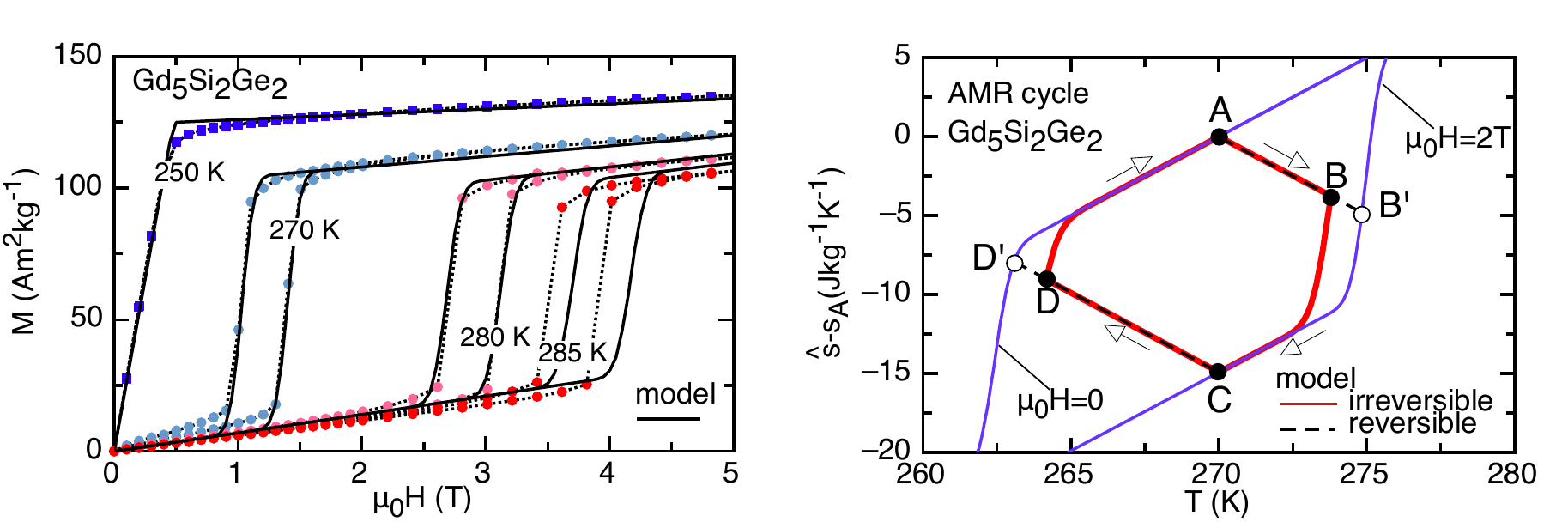}
\caption{Modeling Gd$_5$Si$_2$Ge$_2$. Left: comparison of experimental and modelled $M(H,T)$. Right: model prediction of an active magnetic regenerative refrigeration cycle. After Ref.~\cite{Basso-2007b}}
\label{FIG:GdSiGe}
\end{figure}

The out-of-equilibrium thermodynamics of the system is derived by starting from the assumption that the non-equilibrium Gibbs free energy $g(H,T,b(g_c))$ of the system is a function of the intensive variables $H$ and $T$ and of the internal variable, the function $b(g_c)$. Its expression is given by the superposition of the bistable contributions (Fig.\ref{FIG:H}b):

\beq
g(H,T,b(g_c)) = a(H,T) + \int_{0}^{\infty} dg_{c} \left[ \int_{-\infty}^{b(g_{c})} (g_u - z) p(g_{c}, g_{u}) \, dg_{u} - \int_{b(g_{c})}^{\infty} (g_u - z) p(g_{c}, g_{u}) \, dg_{u} \right] \, ,
\label{EQ: g def}
\eeq

\noindent where $a(H,T)$ is the half sum $a(H,T) = (g_{LT}+g_{HT})/2$. The phase fraction per unit mass $x$ of HT phase is given by:

\beq
x = \int_{0}^{\infty} dg_{c} \int_{-\infty}^{b(g_{c})} \, p(g_{c}, g_{u}) \,dg_{u} \, .
\label{EQ: X def}
\eeq

\noindent The previous expression corresponds to the the Preisach model integral with $z$ as input variable and the phase fraction $x$ as output variable. For a description of the thermodynamic state of the system the aforementioned state-line $b(g_c)$ takes the role of an internal thermodynamic variable not explicitly coupled to intensive variables~\cite{Bertotti-2006}. We make then use of the results known for thermodynamics with internal variables. The extensive variables, magnetization $M$ and specific entropy $s$, are given by the expressions:

\beq
M = - \left. \frac{\partial g }{\partial H} \right|_{T,b(g_c)}
\label{EQ: M def}
\eeq
and
\beq
s = -  \left. \frac{\partial g} {\partial T} \right|_{H,b(g_c)}
\label{EQ: s def}
\eeq

\noindent where the internal variable, the function $b(g_c)$, is kept constant. The rate of entropy production ${d_i \hat{s}}/{dt}$ is given by

\beq
T \frac{d_i s}{dt} = -  \int_{0}^{\infty}  \left. \frac{\delta g} {\delta b(g_c)} \right|_{H,T} \frac{\partial b}{\partial t} \,\,\, dg_c
\label{EQ: Tdis def}
\eeq

\noindent where we have made use of the function derivative. By the fact that the system is not in the equilibrium state every transformation with a change in the state line corresponds to an internal generation of entropy. This is the original and non obvious result obtained by the use of the internal variable thermodynamics. By taking the distribution $p(g_c,g_u)$ independent of $H$ and $T$ the previous expressions are easily computed, giving: 

\beq
M = x M_{HT} +(1-x) M_{LT}
\label{EQ: M X}
\eeq
and
\beq
s = x s_{HT} +(1-x) s_{LT}
\label{EQ: s X}
\eeq

\noindent where $M_{HT} = -\partial g_{HT} / \partial H$, $M_{LT} = -\partial g_{LT} / \partial H$, $s_{HT} = -\partial g_{HT} / \partial T$, $s_{LT} = -\partial g_{LT} / \partial T$ and $x$ is the phase fraction given by the Preisach model expression Eq.(\ref{EQ: X def}) with $z(H,T)$ as input. The rate of entropy production ${d_i \hat{s}}/{dt}$ is thus given by:

\beq
T \frac{d_i s}{dt}  = 2 \int_{0}^{\infty}  [z-b(g_c)] \, p(g_c,b(g_c)) \, \frac{\partial b}{\partial t} \,\,\, dg_c \, .
\label{EQ: Tdis}
\eeq

\noindent The previous expressions can be easily computed by analytic or numerical means once the Preisach distribution $p(g_c,g_u)$ and the Gibbs free energies of the pure phases, $g_{LT}(H,T)$ and $g_{HT}(H,T)$ are known. The model described here has been applied to magnetocaloric materials with hysteresis in order to show how it can solve the problems related to the application of the Maxwell relations to hysteresis curves (see Fig.~\ref{FIG:Maxwell})~\cite{Basso-2006, Basso-2007, Basso-2008c} and to predict thermodynamic cycles (see Fig.\ref{FIG:GdSiGe})~\cite{Sasso-2005, Basso-2006b, Basso-2007b}.


\end{document}